\documentclass{article}

\makeatletter

\mathchardef\bckslash "075C

\def\bksl;{{\ifmmode \bckslash \else $\bckslash$\relax \fi}}
\def\pmS{\setbox0\hbox{\S}\copy0\kern-.95\wd0\copy0\kern-.92\wd0\box0\relax}
\def\br>#1\endbr>{#1}
\newread\filecheck
\newif\if@nosecno 

\newcount\figreduc 
\newif\ifcapfilemiss
\newif\iffignotfound
\def\placefigure#1{\edef\figid{#1}\relax      
        \capfilemisstrue                      
        \openin\filecheck=\jobname.fig\relax
        \ifeof\filecheck \begin{center}
                         Figure #1 to go here. Caption file missing.\par
                         \end{center}
                         \def\next##1!{\relax}\relax
        \else \closein\filecheck \capfilemissfalse\relax \def\next##1!{\input ##1\relax}\relax \fi
        \fignotfoundtrue
        \next\jobname.fig!\relax 
        \ifcapfilemiss \let\fnext\relax \else
           \iffignotfound \begin{center}
                          Figure #1 not found in caption file.\par
                          \end{center}
                          \let\fnext\relax
           \else\let\fnext\insertfigure \fi \fi
         \fnext}
\long\def\fig#1. #2\endfig{
         \expandafter \splitfigspecs #1[100]x!!\relax
         \ifx\figname\figid \endinput \long\def\figcapt{#2}\relax
                            \fignotfoundfalse
         \fi}
\def\insertfigure{\begin{figure}
                    \dumpfigure{\figname}{\figreduc}\relax
                    \caption{\figcapt}
                  \end{figure}}
\def\splitfigspecs #1[#2]#3!!{\edef\figname{#1}\figreduc=#2\relax}
\def\dumpfigure#1#2{\getfig{#1}\relax
       \ifdim\fight=-100cm \centerline{.EPS file not found for fig. ``#1''}\relax
       \else
       \divide\figwd by 10 \multiply\figwd by #2\relax \divide\figwd by 10
       \divide\fight by 10 \multiply\fight by #2\relax \divide\fight by 10
       \hbox to \textwidth{\dimen0=\hsize \advance\dimen0 by -\figwd
                      \divide\dimen0 by 2
                      \hbox to \dimen0{\hfil}\relax
                      \vbox to \fight{\parindent0pt\hsize=0pt
                               \preglue\putfig{#1}{#2}\postglue}\hfill}\relax \fi}
\def\preglue{\vss}\def\postglue{\relax}

\newif\ifepsfbbfound   
\newif\ifepsffileok    
\newif\ifepsfverbose   
\epsfverbosetrue
\newif\ifpagefig       
\newcount\figreduc     
\newcount\xll          
\newdimen\dxll         
\newcount\yll          
\newdimen\dyll         
\newcount\xur          
\newdimen\dxur         
\newcount\yur          
\newdimen\dyur         
\newdimen\fight        
\newdimen\figwd        

\def\getfig#1{\global\epsfbbfoundfalse
    \openin\filecheck=#1.eps\relax 
    \ifeof\filecheck \global\fight=-100cm\figwd=0pt
    \else
   {\epsffileoktrue \chardef\other=12
    \def\do##1{\catcode`##1=\other}\dospecials \catcode`\ =10
    \catcode`\A=9 \catcode`\^^Z=9
    \catcode`\^^L=9 \catcode`\^^?=9
    \loop
    \read \filecheck to \epsffileline
    {\ifx\epsffileline\empty \gdef\epsffileline{!!}\relax \fi}\relax
    \ifeof\filecheck \epsffileokfalse
    \else  
    \expandafter\epsfaux\epsffileline:. \\%
    \fi
    \ifepsffileok 
     \repeat
   \ifepsfbbfound
   \else
    \ifepsfverbose \message{no bounding box found in file #1.eps }\relax
                  \global\fight=100cm \figwd=100cm
    \fi
   \fi
   }\closein\filecheck
   \fi}%
{\catcode`\%=12 \global\let\epsfpercent=
\long\def\epsfaux#1#2:#3\\{\relax
   \ifx#1\epsfpercent
   {\def\testit{#2}\ifx\testit\epsfbblit
      \expandafter\epsfgrab .#3\\%
      \global\epsffileokfalse
      \global\epsfbbfoundtrue
    \else \ifx\testit\epsftitleline
      \message{title line reached: #2}\global\epsfileokfalse \fi \fi}\relax
   \else
   \def\test{!}\def\wat{#1}\ifx\test\wat
   \else \message{.}\relax
   \fi \fi}%
\def\epsfgrab . #1 #2 #3 #4:.#5\\{%
       \message{1: >#1<}\message{2: >#2<}\message{3: >#3<}\message{4: >#4<}\relax
       \global\xll=#1 \global\dxll=#1bp
       \global\yll=#2 \global\dyll=#2bp
       \global\xur=#3 \global\dxur=#3bp
       \global\yur=#4 \global\dyur=#4bp
       \global\fight=#4bp \global\advance\fight by -#2bp
       \global\figwd=#3bp \global\advance\figwd by -#1bp }

\newcount\figstylenumber
\def\ds@nopsdriver{\let\postglue\vss
        \def\putfig##1##2{\hbox to\figwd{\hss Figure ##1 at
                          \number##2\% goes here.\hss}}}

\def\ds@PSfile{\setfigstyle{1}}
\def\ds@plotfile{\setfigstyle{2}}
\def\ds@include{\setfigstyle{3}}
\def\ds@pstext{\setfigstyle{4}}
\def\ds@psfile{\setfigstyle{5}}
\def\ds@dvitops{\setfigstyle{6}}
\def\ds@epsfile{\setfigstyle{7}}
\def\ds@illustration{\setfigstyle{8}}
\def\ds@dvialw{\setfigstyle{9}}

\def\setfigstyle#1{\ifcase #1
  \setstylezero
  \or \setstyleone
  \or \setstyletwo
  \or \setstylethree
  \or \setstylefour
  \or \setstylefive
  \or \setstylesix
  \or \setstyleseven
  \or \setstyleeight
  \or \setstylenine
  \else \message{Style number too high}\relax
  \fi}

\newcount\fscale

\def\putfig#1#2{\relax
     \fscale=#2\relax  \multiply \fscale by 10
     \special{ps: epsfile #1.eps \number\fscale}}

\def\setstylezero{\def\putfig##1##2{\begin{center}Unknown
     specification of insertion scheme in \bksl;figurestyle
     command\end{center}}}

\def\setstyleone{\let\preglue\vss \let\postglue\relax
       \def\putfig##1##2{\relax
           \fscale=\xur \advance\fscale by -\xll
           \multiply \fscale by \figreduc \divide\fscale by 10
           \putspecialone{##1}{\the\xll}{\the\yll}{\the\xur}{\the\yur}{\the\fscale}}}

\def\putspecialone#1#2#3#4#5#6{\includegraphics{#1.eps}}

\def\setstyletwo{\let\preglue\relax \let\postglue\vss
        \def\putfig##1##2{\relax
            \putspecialtwo{\number\figwd}{\number\fight}{\number\dxll}{\number\dyll}{\number\dxur}{\number\dyur}%
            \special{ps: plotfile ##1.eps}%
            \special{ps::[end] endTexFig}}}

\def\putspecialtwo#1#2#3#4#5#6{\special{ps::[begin] #1 #2 #3 #4 #5 #6 startTexFig}}

\def\setstylethree{\let\preglue\relax \let\postglue\vss
        \def\putfig##1##2{\relax
            \special{ps: psfiginit}
            \putspecialthree{\number\figwd}{\number\fight}{\number\dxll}{\number\dyll}{\number\dxur}{\number\dyur}%
            \special{ps: include ##1.eps}%
            \special{ps: literal "endTexFig "}}}

\def\putspecialthree#1#2#3#4#5#6{\special{ps: literal #1 #2 #3 #4 #5 #6 startTexFig}}

\def\setstylefour{\let\preglue\relax \let\postglue\vss
        \def\putfig##1##2{\relax
            \putspecialfour{\number\figwd}{\number\fight}{\number\dxll}{\number\dyll}{\number\dxur}{\number\dyur}%
            \includegraphics{##1.eps}%
            \special{pstext=endTexFig}}}

\def\putspecialfour#1#2#3#4#5#6{\special{pstext="#1 #2 #3 #4 #5 #6 startTexFig"}}

\def\setstylefive{\let\preglue\vss \let\postglue\relax
        \def\putfig##1##2{\relax
            \divide\xll by 10 \multiply\xll by \figreduc \divide\xll by 10
            \divide\yll by 10 \multiply\yll by \figreduc \divide\yll by 10
            \putspecialfive{##1}{\the\figreduc}{\the\xll}{\the\yll}}}

\def\putspecialfive#1#2#3#4{%
     \includegraphics{#1.eps}}

\def\setstylesix{\let\preglue\vss \let\postglue\relax
        \def\putfig##1##2{\relax
            \putspecialsix{##1}{\number\figwd}{\number\fight}}}

\def\putspecialsix#1#2#3{%
     \special{dvitops: import #1.eps #2sp #3sp}}

\def\setstyleseven{\relax}

\def\setstyleeight{\let\preglue\vss \let\postglue\relax
     \def\putfig##1##2{\relax
     \fscale=##2\relax  \multiply \fscale by 10
     \special{illustration ##1.eps scaled \number\fscale}}}

\def\setstylenine{\let\preglue\relax \let\postglue\vss
        \def\putfig##1##2{\putspecialnine{##1}{\number\figreduc}}}

\def\putspecialnine#1#2{%
     \special{language "PS", literal "0.#2 0.#2 scale", include "#1.eps"}}

\newif\iftabfilemiss
\def\column#1{\hbox to #1{\hfill}}
\def\placetable#1{\tabfilemisstrue         
        \openin\filecheck=\jobname.t#1 \relax
        \ifeof\filecheck \begin{center}
                          Table #1 to go here. Table file missing.\par
                          \end{center}\relax
                          \tabfilemisstrue
           \else \closein\filecheck \tabfilemissfalse \relax
        \fi
        \iftabfilemiss \let\tnext\@gobble \else \let\tnext\inserttable \fi
        \tnext{#1}\relax}
\def\inserttable#1{\begin{table}
                   \small
                   \input \jobname.t#1\relax
                   \end{table}}
\def\Table #1. #2\par{\begin{centering}{Table #1}\\#2\par
                      \end{centering}}

\def\@procti{}
\def\@proctI#1{#1}
\def\@proctII#1{\uppercase\expandafter{#1}}
\def\@proctIII#1{#1}
\def\@setprocti#1{\ifcase #1
                  \or \let\@procti\@proctI
                  \or \let\@procti\@proctII
                  \or \let\@procti\@proctIII
                  \else \let@procti
                  \fi}

\def\@startsection#1#2#3#4#5#6{\if@noskipsec \leavevmode \fi
   \@setprocti{#2}\relax
   \par \@tempskipa #4\relax
   \@afterindenttrue
   \ifdim \@tempskipa <\z@ \@tempskipa -\@tempskipa \@afterindentfalse\fi
   \if@nobreak \everypar{}\else
     \addpenalty{\@secpenalty}\addvspace{\@tempskipa}\fi \@ifstar
     {\@ssect{#3}{#4}{#5}{#6}}{\@dblarg{\@sect{#1}{#2}{#3}{#4}{#5}{#6}}}}

\def\@sect#1#2#3#4#5#6[#7]#8{%
     \ifnum #2>\c@secnumdepth \let\@svsec\@empty
        \else \def\leeg{}\relax
        \ifx\leeg\thesection \let\@svsec\@empty
           \else \refstepcounter{#1}%
                 \edef\@svsec{\csname the#1\endcsname.\hskip 1em}\relax
        \fi
     \fi
     \@tempskipa #5\relax
      \ifdim \@tempskipa>\z@
        \begingroup #6\relax
          \@hangfrom{\hskip #3\relax\@svsec}{\interlinepenalty \@M \@procti{#8}\checkheader{#7}\par}%
        \endgroup
       \csname #1mark\endcsname{#7}\addcontentsline
         {toc}{#1}{\if@nosecno
                      \else
                      \ifnum #2>\c@secnumdepth
                         \else \protect\numberline{\csname the#1\endcsname}\fi
                   \fi
                    #8}\else
        \def\@svsechd{#6\hskip #3\relax  
                   \@svsec #8\csname #1mark\endcsname
                      {#7}\addcontentsline
                           {toc}{#1}{\ifnum #2>\c@secnumdepth \else
                             \protect\numberline{\csname the#1\endcsname}\fi
                       \@procti{#8}\checkheader{#7}}}\fi
     \@xsect{#5}}

\def\@xsect#1{\@tempskipa #1\relax
      \ifdim \@tempskipa>\z@
       \par \nobreak
       \vskip \@tempskipa
       \@afterheading
    \else \global\@nobreakfalse \global\@noskipsectrue
       \everypar{\if@noskipsec \global\@noskipsecfalse
                   \clubpenalty\@M \hskip -\parindent
                   \begingroup \@svsechd \endgroup \unskip
                   \hskip -#1\relax  
                  \else \clubpenalty \@clubpenalty
                    \everypar{}\fi}\fi\ignorespaces}

\def\@ssect#1#2#3#4#5{\@tempskipa #3\relax
   \ifdim \@tempskipa>\z@
     \begingroup #4\@hangfrom{\hskip #1}{\interlinepenalty \@M \@procti{#5}\par}\endgroup
   \else \def\@svsechd{#4\hskip #1\relax #5}\fi
    \@xsect{#3}}

\begin{document}
\@addtoreset{equation}{section}
\def\theequation{\thesection.\arabic{equation}}

\def\appendix{\vskip1pc\par {\bf Appendix} \gdef\pmS{}\relax
  \setcounter{section}{0}
  \setcounter{subsection}{0}
  \def\thesection{\Alph{section}}}


\evensidemargin=\oddsidemargin


\newif\ifwidehead
\def\markright#1{%
     \ifwidehead \gdef\rightmarker{Long header; use \bksl;section[short header]$\{$title$\}$}\relax
        \else \gdef\rightmarker{#1}\relax
     \fi
     {\let\protect\noexpand
     \let\label\relax \let\index\relax \let\glossary\relax
     \expandafter\@markright\@themark
     {\rightmarker}\mark{\@themark}}\if@nobreak\ifvmode\nobreak\fi\fi}

\def\markleft#1{\checkheader{#1}\relax
     \ifwidehead \gdef\leftmarker{Heading too long! define short one}\relax
        \else \gdef\leftmarker{#1}\relax
     \fi
     {\let\protect\noexpand
     \let\label\relax \let\index\relax \let\glossary\relax
     \expandafter\@markleft\@themark
     {\leftmarker}\mark{\@themark}}\if@nobreak\ifvmode\nobreak\fi\fi}

\def\@markleft#1#2#3{\gdef\@themark{{#3}{#2}}}

%
\def\@themark{{}{}}
\gdef\leftmarker{} \gdef\specleftmarker{} \gdef\rightmarker{}


\newif\if@shorttitle
\def\shorttitle#1{\global\@shorttitletrue
                  \gdef\specleftmarker{#1}}

\def\checkheader#1{\setbox0\hbox{\small\uppercase\expandafter{#1}}\relax
       \ifdim\wd0<\headroom \global\wideheadfalse
          \else \global\wideheadtrue \ifvmode \leavevmode \fi
       \fi
       \ifwidehead \marginpar{\tiny\bf
                    Long heading! Define short title}\relax
       \fi}


\def\@maketitle{\newpage
                \thispagestyle{plain}
                \begin{centering}
PROGRESS IN OPTICS XXXVI, pp. 245-294 \\
                E. WOLF, Editor, Elsevier, Amsterdam, 1996\\

                \gdef\chapno{X}\relax           
                \vskip 5\baselineskip
                {\bf \chapno}
                \vskip 2\baselineskip
                \markleft{FULL}\relax
                \wideheadfalse
                \noindent {\bf \@title}\relax
                \if@shorttitle \markleft{\specleftmarker}\relax \fi
                \checkheader{\leftmarker}\relax
                {\ifwidehead \markleft{Please specify shorter title!}\relax \fi}\relax
                \par
                \vskip 2\baselineskip
                {\small BY}\par
                \vskip 2\baselineskip
                \relax \null
                {\large \sc
                  \@author\par}
                \end{centering}}

\def\and{\par\vskip 1pc}


\def\tableofcontents{{\vfill\eject \relax
                 \pagestyle{plain}
                 \putstartcon
                 \@starttoc{toc}\relax
                 \putendcon
                 \vfill\eject}\relax
                 \pagestyle{headings}}


\def\putstartcon{\vbox to 6\baselineskip{\vss \centerline{\bf CONTENTS}}\relax
                 \vskip 4\baselineskip
                 \rightline{\small PAGE}
                 \vskip0pt
                 \begingroup
                 \relax}
\def\putendcon{\endgroup}

\def\numberline#1{\setbox0\hbox{#1}\relax
                  \ifdim\wd0>\@tempdima \box0\enspace
                     \else \hbox to\@tempdima{#1\hfil}\relax
                  \fi}



\def\period{}

\def\ps@headings{\let\@mkboth\markboth
 \def\@oddfoot{}\def\@evenfoot{}
  \def\@evenhead{ \rm \thepage\hfil {\def\period{. }\small \leftmark}}
  \def\@oddhead{{\def\period{. }\small \rightmark}\hfil \rm\thepage}
  \def\sectionmark##1{\markright {\ifx\thesection\@empty \else
     \ifnum \c@secnumdepth >\z@ \thesection\protect\period\relax \fi\fi
     \uppercase\expandafter{##1}}{}}%
  \def\subsectionmark##1{\relax}}

\def\thesection       {\arabic{section}}
\def\thesubsection    {\thesection.\arabic{subsection}}
\def\thesubsubsection {\thesubsection .\arabic{subsubsection}}
\def\theparagraph     {\thesubsubsection.\arabic{paragraph}}
\def\thesubparagraph  {\theparagraph.\arabic{subparagraph}}
\def\theequation      {\thesection.\arabic{equation}}

\def\section{\@startsection {section}{1}{\z@}{-3.5ex plus-1ex minus
    -.2ex}{2.3ex plus.2ex}{\reset@font\normalsize \leavevmode \pmS\,\bf}}
\def\subsection{\@startsection{subsection}{2}{\z@}{-3.25ex plus-1ex
    minus-.2ex}{1.5ex plus.2ex}{\reset@font\normalsize\rm}}
\def\subsubsection{\@startsection{subsubsection}{3}{\z@}{-3.25ex plus
    -1ex minus-.2ex}{1.5ex plus.2ex}{\reset@font\normalsize}}
\def\paragraph{\@startsection
     {paragraph}{4}{\z@}{3.25ex plus1ex minus.2ex}{-1em}{\reset@font
     \normalsize\rm}}
\def\subparagraph{\@startsection
     {subparagraph}{4}{\parindent}{3.25ex plus1ex minus
     .2ex}{-1em}{\reset@font\normalsize\rm}}

\def\pagebox#1{\hbox to 2pc{#1}}
\newdimen\headroom
\headroom=\hsize             
\setbox0\pagebox{1\hss}\relax \advance\headroom by -2\wd0

\ds@twoside


      \def\thebibliography#1{\@nosecnotrue
      \gdef\thesection{}\relax
      \section{\refname}\relax
      \begingroup
       \tolerance 9999
       \hbadness 9999
       \widowpenalty 10000
       \clubpenalty 200
       \small}

\def\endthebibliography{\endgroup
  \def\@noitemerr{\@warning{Empty `thebibliography' environment}}%
  \endlist}

\def\bibitem{\@ifnextchar[{\@lbibitem}{\@bibitem}}

\def\@lbibitem[#1]#2{\@putbiblabel{#1}\if@filesw
      {\def\protect##1{\string ##1\space}\immediate
       \write\@auxout{\string\bibcite{#2}{#1}}}\fi\ignorespaces}

\def\@putbiblabel#1{\par\noindent #1\par\nobreak\noindent \hangindent=1em
                    \hangafter=0 \relax}

\def\@bibitem#1{\item\if@filesw \immediate\write\@auxout
       {\string\bibcite{#1}{\the\value{\@listctr}}}\fi\ignorespaces}

\def\bibcite#1#2{\global\@namedef{b@#1}{#2}}

\let\citation\@gobble

\def\cite{\@ifnextchar [{\@tempswatrue\@citex}{\@tempswafalse\@citex[]}}


\def\@citex[#1]#2{\if@filesw\immediate\write\@auxout{\string\citation{#2}}\fi
  \let\@citea\@empty
  \@cite{\@for\@citeb:=#2\do
    {\@citea\def\@citea{, }%
     \def\@tempa##1##2\@nil{\edef\@citeb{\if##1\space##2\else##1##2\fi}}%
     \expandafter\@tempa\@citeb\@nil
     \@ifundefined{b@\@citeb}{{\reset@font\bf ?}\@warning
       {Citation `\@citeb' on page \thepage \space undefined}}%
     {\csname b@\@citeb\endcsname}}}{#1}}

\let\bibdata=\@gobble
\let\bibstyle=\@gobble

\def\bibliography#1{\if@filesw\immediate\write\@auxout{\string\bibdata{#1}}\fi
  \@input{\jobname.bbl}}

\def\bibliographystyle#1{\if@filesw\immediate\write\@auxout
    {\string\bibstyle{#1}}\fi}

\def\nocite#1{\@bsphack
  \if@filesw\immediate\write\@auxout{\string\citation{#1}}\fi
  \@esphack}


\def\@cite#1#2{{#1\if@tempswa , #2\fi}}
\def\@biblabel#1{[#1]}

\def\@listI{\leftmargin\leftmargini \parsep 0\p@ plus1\p@ minus\p@
\topsep 0\p@ plus3\p@ minus\p@
\itemsep 0\p@ plus1\p@ minus\p@}



\@ifundefined{ttfam}{\let\bckslash\backslash}{\mathchardef\bckslash "075C}
\@ifundefined{bi}{\def\bi{\bf}}{\relax} \@ifundefined{tt}{\let\tt\rm}{} \ps@headings

\textheight=39.25pc \textwidth=27pc \headsep 12pt \topskip = 10pt \footskip 12pt \marginparwidth 48pt

\widowpenalty 10000
\parindent1em

\title{PHOTON WAVE FUNCTION}

\author{Iwo Bialynicki-Birula\\
{\em Center for Theoretical Physics, Polish Academy of Sciences\\ Al.
Lotnik\'ow 32, 02-668 Warsaw, Poland\\and\\
Rochester Theory Center for Optical Science and Engineering\\
University of Rochester, Rochester, NY 14627, USA }}
\maketitle

\tableofcontents
\section{Introduction}

Photon wave function is a controversial concept. Controversies stem from the
fact that photon wave functions can not have all the properties of the
Schr\"odinger wave functions of nonrelativistic wave mechanics. Insistence
on those properties that, owing to peculiarities of photon dynamics, cannot
be rendered, led some physicists to the extreme opinion that the photon wave
function does not exist. I reject such a fundamentalist point of view in
favor of a more pragmatic approach. In my view, the photon wave function
exists as long as it can be precisely defined and made useful. Many authors
whose papers are quoted in this review share the same opinion and had no
reservations about using the name the photon wave function when referring to
a complex vector-function of space coordinates ${\bf r}$ and time $t$ that
adequately describes the quantum state of a single photon.

The notion of the photon wave function is certainly not new, but strangely
enough it has never been systematically and fully explored. Some textbooks
on quantum mechanics start the introduction to quantum theory with a
discussion of photon polarization measurements (cf., for example
\cite{Dirac_58}, \cite{Baym_69}, \cite{Lipkin_73}, \cite{CTDL_77}), but in
all these expositions a complete photon wave function never takes on a
specific mathematical form. Even Dirac who writes ``The essential point is
the association of each of the translational states of the photon with one
of the wave functions of ordinary wave optics'', never expresses  this association in an explicit form. In this context he also uses the now famous
phrase: ``Each photon interferes only with itself'' which implies the
existence of photon wave functions whose superposition leads to interference
phenomena.

In the textbook analysis of polarization, only simple prototype
two-com\-po\-nent wave functions are used to describe various polarization
states of the photon and with their help the preparation and the measurement
of polarization is analyzed. However, it is not explained, why a wave
function should not be used to describe also the "translational states of
the photon" mentioned by Dirac. After such a heuristic introduction to
quantum theory, the authors go on to the study of massive particles and if
they ever return to quantum theory of photons it is always within the
formalism of second quantization with creation and annihilation operators.
In some textbooks (cf., for example, \cite{Bohm_54}, \cite{Power_64}) one
may even find statements that completely negate the possibility of
introducing a wave function for the photon.

A study of the photon wave function should be preceded by an explanation
what is the photon and why a description of the photon in terms of a wave
function must exist. According to modern quantum field theory, photons,
together with all other particles (and also quasiparticles, phonons,
excitons, plasmons, etc.), are the {\em quantum excitations} of a field. In
the case of photons, these are the excitations of the electromagnetic field.
The lowest field excitation of a {\em given type} corresponds to one photon and
higher field excitations involve more than one photon. This concept of a
photon (called the modern photon in a tutorial review by \cite{KAA_89})
enables one to use the photon wave function not only to describe quantum
states of an excitation of the free field but also of the electromagnetic
field interacting with a medium. Conceptually, the difference between free
space and a medium is not essential since the physical vacuum is like a
polarizable medium. It is filled with all the virtual pairs --- zero point excitations of charged quantum fields. Therefore, even in free space, photons can be also viewed as the excitations of the vacuum made mostly of virtual
electron-positron pairs (\cite{IBB_63}, \cite{Bjorken_63}).

Even though, in principle, all particles can be treated as field
excitations, photons are much different from massive particles. They are
also different from massless neutrinos since the photon number does not obey
a conservation law. There are problems with the photon localization and as a
result the position operator for the photon is ill-defined, but the
similarities between photons and other quantum particles are so ample that
the introduction of the photon wave function seems to be fully justified and
even necessary in order to achieve a complete unification of our description
of all particles.

\subsection{COORDINATE VS. MOMENTUM REPRESENTATION}

In nonrelativistic quantum mechanics the term {\em coordinate
representation} is used to denote the representation in which the wave
function $\psi({\bf r})$ is defined as a projection of the state vector
$\vert\psi\rangle$ on the eigenstates $\vert{\bf r}\rangle$ of the
components $\hat{x}, \hat{y}$, and $\hat{z}$ of the position operator $\hat{\bf r}$,
\begin{eqnarray}
\psi({\bf r}) = \langle{\bf r}\vert\psi\rangle.\label{wf_pos}
\end{eqnarray}
The wave function in coordinate representation, therefore, becomes
automatically a function of the eigenvalues of the position operator $\hat{\bf
r}$. The position operators act on the wave function simply through a
multiplication. In quantum mechanics of photons this approach does not work
due to difficulties with the definition of the photon position operator (cf.
\S \ref{localizability}). One may still, however, introduce functions of the
coordinate vector ${\bf r}$ to describe quantum states of the photon. By
adopting this less stringent point of view that does not tie the wave
function in coordinate representation with the formula (\ref {wf_pos}), one
avoids the consequences of the nonexistence of the photon position operator
$\hat{\bf r}$. In principle, any function of ${\bf r}$ that adequately describes
photon states may be called a photon wave function in coordinate
representation and it is a matter of taste and convenience which one to use.
It should be pointed out that in a relativistic quantum theory, even for
particles with nonvanishing rest mass, the position operator and the
localization associated with it do not live up to our nonrelativistic
expectations. The differences in localization of photons and, say electrons,
are more quantitative then qualitative since they amount to the "spilling of
the wave function" beyond the localization region governed by a power law
versus an exponential decay.

The photon wave function in {\em momentum representation} has not stirred
any controversy since the photon momentum operator $\hat{\bf p}$ is well defined. Its
existence, as the generator of translations, follows directly from the
general theory of representations of the Poincar\'e group developed by
\cite{Wigner_39}. It has always been taken for granted by all physicists
working in relativistic quantum electrodynamics that the notion of the
photon wave function in momentum representation is well founded. Such wave
functions describing initial and final states of photons appear in all
formulas for transition amplitudes in the S-matrix theory of scattering
phenomena (cf., for example, \cite{Schweber_61}, \cite{AB_65},
\cite{BBBB_75}, \cite{CDG_89}). Thus, one may safely assert that the photon
wave function in momentum representation is a well defined and fully
established object.

\subsection{PHASE REPRESENTATION}

The photon wave functions discussed in this review are distinct from the
{\em one-mode} wave functions that have been introduced in the past
(\cite{London_27}, \cite{BBBB_76}, \cite{PB_88}) to describe {\em multi-photon
states}. These functions depend on the phase $\varphi$ of the field and were
called the wave functions in the phase representation by \cite{BBBB_76}. The
wave functions $\Psi(\varphi)$ characterize quantum states of a {\em
selected mode} of the quantized electromagnetic field and, in general, they
describe a superposition of states with different numbers of photons. All
spatial characteristics of these states are contained in the mode function
that defines the selected mode of the electromagnetic field. One-mode wave
functions $\Psi(\varphi)$ describe properties of multi-photon states of the
quantized electromagnetic field with all photons being in the same quantum mechanical state. In contrast, the photon wave function in the coordinate representation can be
identified with the mode function itself (cf. \S \ref{mode}). It describes a
state of a {\em single} photon and not a state of the quantized field.

\subsection{LANDAU-PEIERLS WAVE FUNCTION}

The concept of the photon wave function in coordinate representation was
introduced for the first time by \cite{LP_30}. The same function has been
independently rediscovered more recently by \cite{Cook_82a, Cook_82b}, and
\cite{Inagaki_94}. The Landau-Peierls proposal has not been met with great
enthusiasm since their wave function is a highly nonlocal object.

The nonlocality of the Landau-Peierls wave function is introduced by
operating on the local electromagnetic field with the integral operator
$(-\Delta)^{-1/4}$,
\begin{eqnarray}
 ((-\Delta)^{-1/4}f)({\bf r}) = \pi\int \frac{d^3r'}
 {(2\pi\vert  {\bf r} - {\bf r}'\vert)^{5/2}}f({\bf r}').\label{lp_loc}
\end{eqnarray}
This integral operator corresponds to a division by $\sqrt{\vert{\bf
k}\vert}$ of the Fourier transform and it changes the dimension of the wave
function from $L^{-2}$, characteristic of the electromagnetic field, to
$L^{-3/2}$. Therefore the modulus squared of the Landau-Peierls wave
function has the right dimensionality to be interpreted as a probability
density to find a photon. In particular, these wave functions can be
normalized to one with the standard definition of the norm since the
integral of the modulus squared of the wave function is dimensionless.
However, as has been already noted by \cite{Pauli_33}, despite its right
dimensionality the nonlocal wave function has serious drawbacks. First, it
does not transform under Lorentz transformations as a tensor field or any
other geometric object. Second, a nonlocal wave function taken at a point in
one coordinate system depends on the values of this wave function in {\em
all of space} in another coordinate system. Third, the probability density
defined with the use of a nonlocal wave function does not correspond to the
probability of interaction of localized charges with the electromagnetic
field. The vanishing of the wave function at a definite point, in Pauli's
words (\cite{Pauli_33}), has ``no direct physical significance'' because the
electromagnetic field does act on charges at the points where the
probability to find a photon vanishes. The Landau-Peierls wave functions can
not be used as primary objects in the presence of a medium since one is
unable to impose proper boundary conditions on such nonlocal objects. These
functions can be introduced, if one wishes so, as secondary objects related
to the local wave function by a nonlocal transformation (\S
\ref{interpretation}). Scalar products and expectation values look simpler
when they are expressed in terms of the Landau-Peierls wave functions but
that is, perhaps, their only advantage.

\subsection{RIEMANN-SILBERSTEIN WAVE FUNCTION}

The mathematical object that fully deserves the name of the photon wave
function can be traced back to a complexified form of Maxwell's equations
that was known already at the turn of the century. The earliest reference is
the second volume of the lecture notes on the differential equations of
mathematical physics by Riemann that were edited and published by
\cite{Weber_01}. Various applications of this form of Maxwell's equations
were given by \cite{Silberstein_07a}, \cite{Silberstein_07b},
\cite{Silberstein_14} and \cite{Bateman_15} in the framework of classical
physics.

The complex form of Maxwell's equations is obtained by multiplying the first
pair of these equations
\begin{eqnarray}
\partial_t {\bf D}({\bf r},t) = {\bf\nabla}\times{\bf H}({\bf r},t),
\;\;\;\;{\bf\nabla}\!\cdot\!{\bf D}({\bf r},t) = 0,\label{max1}
\end{eqnarray}
by the imaginary unit and then by subtracting from it the second pair
\begin{eqnarray}
\partial_t {\bf B}({\bf r},t) = -{\bf\nabla}\times{\bf E}({\bf r},t),
\;\;\;\;{\bf\nabla}\!\cdot\!{\bf B}({\bf r},t) = 0.\label{max2}
\end{eqnarray}
In the SI units that are used here the vectors ${\bf D}$ and ${\bf B}$ have
different dimensions and prior to subtraction one must equalize the
dimensions of both terms. The resulting equations in empty space are
\begin{eqnarray}
&&i\partial_t{\bf F}({\bf r},t)
 = c\nabla\times{\bf F}({\bf r},t),\label{cmax1}\\
 &&\nabla\!\cdot\!{\bf F}({\bf r},t) = 0,\label{cmax2}
\label{cplxm}
\end{eqnarray}
where
\begin{eqnarray}
 {\bf F}({\bf r},t) = (\frac{{\bf D}({\bf r},t)}{\sqrt{2\epsilon}_0}
 + i\frac{{\bf B}({\bf r},t)}{\sqrt{2\mu}_0}),\label{rsv0}
\end{eqnarray}
and $c = 1/\sqrt{\epsilon_0\mu_0}$. Around the year 1930 Majorana
(unpublished notes quoted by \cite{MRB_74}) arrived at the same complex
vector exploring the analogy between the Dirac equation and the Maxwell
equations. \cite{Kramers_38} made an extensive use of this vector in his
treatment of quantum radiation theory. This vector is also a natural object
to use in the quaternionic formulation of Maxwell's theory
(\cite{Silberstein_14}). With the advent of spinor calculus that superseded
the quaternionic calculus, the transformation properties of the
Riemann-Silber\-stein vector have become even more transparent. When
Maxwell's equations were cast into the spinor form (\cite{LU_31},
\cite{Oppenheimer_31}), this vector turned into a symmetric second-rank
spinor. The use of the Riemann-Silberstein vector as the wave function of
the photon has been advocated by \cite{Oppenheimer_31, Moliere_49, Good_57,
IBB_94, Sipe_95}, and \cite{IBB_96a}.

It has already been noticed by \cite{Silberstein_07a} that the important
dynamical quantities associated with the electromagnetic field: the energy
density and the Poynting vector can be represented as {\em bilinear
expressions} built from the complex vector ${\bf F}$. Using modern terminology, one
would say that the formulas for the energy $E$, momentum ${\bf P}$, angular
momentum ${\bf M}$, and the moment of energy ${\bf N}$ of the
electromagnetic field look like quantum-mechanical expectation values
\begin{eqnarray}
\arraycolsep=2pt
 E &=& \int\!d^3r\,{\bf F}^*\!\cdot\!{\bf F},\label{epm1}\\
 {\bf P} &=& \frac{1}{2ic}\!\int\!d^3r\,{\bf F}^*\!\times\!{\bf F},
 \label{epm2}\\
 {\bf M} &=& \frac{1}{2ic}\!
 \int\!d^3r\,{\bf r}\times({\bf F}^*\!\times\!{\bf F}),\label{epm3}\\
 {\bf N} &=& \int\!d^3r\,{\bf r}\,({\bf F}^*\!\cdot\!{\bf F}),\label{epm4}
 \end{eqnarray}
evaluated in a state described by the wave function ${\bf F}$. All these
quantities are invariant under the multiplication of ${\bf F}$ by a phase
factor $\exp(i\alpha)$. Such a multiplication results in the so called
duality rotation (\cite{MW_57}) of the field vectors
\begin{eqnarray}
 {\bf D}'/\sqrt{\epsilon_0} = \cos\alpha{\bf D}/\sqrt{\epsilon_0}
 - \sin\alpha{\bf B}/\sqrt{\mu_0},\\
 {\bf B}'/\sqrt{\mu_0} = \cos\alpha{\bf B}/\sqrt{\mu_0}
 + \sin\alpha{\bf D}/\sqrt{\epsilon_0}.
\end{eqnarray}
The coupling of the electromagnetic field with charges fixes the phase
$\alpha$ but for a free photon one has the same complete freedom in choosing
the overall phase of the photon wave function as in standard wave mechanics
of massive particles.

The Riemann-Silberstein vector has also many other properties that one would
associate with a one-photon wave function, except for a somewhat modified
probabilistic interpretation. Insistence on exactly the same form of the
expressions for transition probabilities as in nonrelativistic wave
mechanics leads back to the Landau-Peierls wave function with its highly
nonlocal transformation properties.

\section{Wave equation for photons\label{wave_equation}}

The wave equation for the photon is taken to be the complexified form
(\ref{cmax1}) of Maxwell's equations. In order to justify this choice, one
may show (cf. \S \ref{momentum}) that the Fourier decomposition of the
solutions of this wave equation leads to the same photon wave functions in
momentum representation that can be introduced without any reference to a
wave equation directly from the general theory of representations of the Poincar\'e
group (\cite{BW_48}, \cite{LM_62}). There is also a heuristic argument
indicating that eq. (\ref{cmax1}) is the right choice. Namely, the wave
equation (\ref{cmax1}) can be written in the same form as the Weyl equation
for the neutrino wave function. As a matter of fact, all wave equations for
massless particles with arbitrary spin can be cast into the same form (cf. \S
\ref{spinor}).

\subsection[WAVE EQUATION IN FREE SPACE]{WAVE EQUATION FOR PHOTONS IN FREE
SPACE}

In order to see a correspondence between Maxwell's equations and quantum
mechanical wave equations, one may follow \cite{Oppenheimer_31} and
\cite{Moliere_49} and rewrite (\ref{cmax1}) with the use of the spin-1
matrices $s_x, s_y, s_z$ well known from quantum mechanics (see, for
example, \cite{Schiff_68}). The matrices that will be used here are in a
different  representation from the one usually used in quantum mechanics since they
act on the Cartesian vector components of the wave function and not on the
components labeled by the eigenvalues of $s_z$. That is the reason why the
matrix $s_z$ is not diagonal.
\begin{eqnarray}
 s_x = \left[ \begin{array}{ccc}
 0 & 0 & 0\\
 0 & 0 & -i\\
 0 & i & 0\end{array}
 \right]\!,\;
 s_y = \left[ \begin{array}{ccc}
 0 & 0 & i\\
 0 & 0 & 0\\
 -i & 0 & 0\end{array}
 \right]\!,\;
 s_z = \left[ \begin{array}{ccc}
 0 & -i & 0\\
 i & 0 & 0\\
 0 & 0 & 0\end{array}
 \right]\!.\label{spin_mat}
\end{eqnarray}
Eq. (\ref{cmax1}) can be written in terms of these spin matrices if the
following conversion rule from vector notation to matrix notation is applied
\begin {eqnarray}
{\bf a}\times{\bf b} = -i({\bf a}\!\cdot\!{\bf s}){\bf b}.
\end {eqnarray}
The resulting equation
\begin{eqnarray}
 i\hbar\partial_t{\bf F}({\bf r},t) = c\Big({\bf s}\!\cdot\!
 \frac{\hbar}{i}{\bf\nabla}\Big) {\bf F}({\bf r},t),\label{mf}
\end{eqnarray}
is of a Schr\"odinger type but with a different Hamiltonian. The divergence
condition (\ref{cmax2}) can also be expressed in terms of spin matrices
either as
\begin{eqnarray}
 ({\bf s}\!\cdot\!{\bf \nabla}) s_j\,{\bf F}
 = \nabla_j\,{\bf F},\label{aux_s}
\end{eqnarray}
or equivalently (\cite{Pryce_48}) as
\begin{eqnarray}
 \Big({\bf s}\!\cdot\!{\bf\nabla}\Big)^2{\bf F}
 = \Delta{\bf F}.
\end{eqnarray}
The form (\ref{mf}) of the Maxwell equations compares directly with the Weyl
equation for neutrinos (\cite{Weyl_29})
\begin{eqnarray}
 i\hbar\partial_t{\phi}({\bf r},t)
 = c\Big(\mbox{\boldmath$\sigma$}\!\cdot\!
 \frac{\hbar}{i}{\bf\nabla}\Big) {\phi}({\bf r},t).\label{weyl}
\end{eqnarray}
Eq. (\ref{weyl}) differs from (\ref{mf}) only in having the Pauli matrices,
appropriate for spin-1/2 particles, instead of the spin-1 matrices that are
appropriate for photons. Of course, one may cancel the factors of $\hbar$
appearing on both sides of eqs. (\ref{mf}) and (\ref {weyl}), but their
presence makes the connection with quantum mechanics more transparent.

Some authors (\cite{Oppenheimer_31}, \cite{Ohmura_56}, \cite{Moses_59})
introduced a different, though equivalent, form of the photon wave equation
for a {\em four-component} wave function. The inclusion of the forth component
enables one to incorporate the divergence condition in a natural way. This
approach is directly related to the spinorial representation of the photon
wave function and will be discussed in \S \ref{spinor}.

In quantum mechanics the stationary solutions of the wave equation play a
distinguished role. They are the building blocks from which all solutions
can be constructed. Stationary solutions of the wave equation are obtained
by separating the time variable and solving the resulting eigenvalue
problem. The eigenvalue equation resulting from the photon wave equation
(\ref {mf}) is
\begin{eqnarray}
 c\Big({\bf s}\!\cdot\!\frac{\hbar}{i}{\bf \nabla}\Big){\bf F}({\bf r})
 =  \hbar\omega{\bf F}({\bf r}).\label{mfe}
\end{eqnarray}
Assuming that the photon energy $\hbar\omega$ is positive, one reads from
(\ref {mfe}) that the projection of the spin on the direction of momentum
(helicity) is positive. It can easily be checked that one can reverse this
sign by changing $i$ into $-i$ in the definition (\ref {rsv0}) of the
Riemann-Silberstein vector. Thus, the choice of sign in the definition of
this vector is equivalent to choosing positive or negative helicity,
corresponding to left-handed or right-handed circular polarization. In order
to account for both helicity states of the photons, one has to consider both
vectors; one may call them ${\bf F}_+$ and ${\bf F}_-$. This doubling of the
 vectors ${\bf F}$ has already been considered by \cite{Silberstein_14} in
the context of the classical Maxwell equations.

Of course, if one is only interested in translating the Maxwell equations
into a complex form, one can restrict oneself to either ${\bf F}_+$ or ${\bf
F}_-$. In both cases one obtains a one-to-one correspondence between the
real field vectors ${\bf D}$ and ${\bf B}$ and their complex combination
${\bf F}_{\pm}$. However, to have a bona fide photon wave function one must
be able to superpose different helicity states without changing the sign of
the energy (frequency). This can only be done when both helicities are
described by different components of {\em the same wave function} as in the
theory of spin-1/2 particles. One can see even more clearly the need to use both
complex combinations when one deals with the propagation of photons in a
medium.

\subsection[WAVE EQUATION IN A MEDIUM]{WAVE EQUATION FOR PHOTONS IN A
MEDIUM}

In free space the two vectors ${\bf F}_{\pm}$ satisfy two separate wave
equations
\begin{eqnarray}
 i\partial_t{\bf F}_{\pm}({\bf r},t)
 = {\pm}c{\bf\nabla}\times{\bf F}_{\pm}({\bf r},t).
 \label{two_mf}
\end{eqnarray}
In a homogeneous medium, using the values of $\epsilon$ and $\mu$ for the
medium in the definition of the vectors ${\bf F}_{\pm}$,
\begin{eqnarray}
 {\bf F}_{\pm}({\bf r},t) = (\frac{{\bf D}({\bf r},t)}{\sqrt{2\epsilon}}
 \pm i\frac{{\bf B}({\bf r},t)}{\sqrt{2\mu}}),\label{rsv}
\end{eqnarray}
one also obtains two separate wave equations. The new vectors ${\bf
F}_{\pm}$ are linear combinations of the free-space vectors ${\bf
F}^0_{\pm}$,
\begin{eqnarray}
 {\bf F}_+ = \frac{1}{2}\bigl[\bigl(\sqrt{\frac{\epsilon_0}{\epsilon}}
 + \sqrt{\frac{\mu_0}{\mu}}\bigr){\bf F}^0_+
 + \bigl(\sqrt{\frac{\epsilon_0}{\epsilon}}
 - \sqrt{\frac{\mu_0}{\mu}}\bigr){\bf F}^0_-\bigr],\\
 {\bf F}_- = \frac{1}{2}\bigl[\bigl(\sqrt{\frac{\epsilon_0}{\epsilon}}
 - \sqrt{\frac{\mu_0}{\mu}}\bigr){\bf F}^0_+
 + \bigl(\sqrt{\frac{\epsilon_0}{\epsilon}}
 + \sqrt{\frac{\mu_0}{\mu}}\bigr){\bf F}^0_-\bigr].
\end{eqnarray}
Thus, the positive and negative helicity states in a medium are certain
linear superpositions of such states in free space. The necessity to form
linear superpositions of both helicity states shows up even more forcefully in an inhomogeneous medium, because then it is not possible to split the wave
equations into two independent sets. For a linear, time-independent,
isotropic medium, characterized by space-dependent permittivity and
permeability, one obtains the following coupled set of wave equations
\begin{eqnarray}
 i\partial_t{\bf F}_+({\bf r},t)
 &=& v({\bf r})\bigl({\bf\nabla}\times{\bf F}_+({\bf r},t)\label{inh1}\\
 &-&\frac{1}{2v({\bf r})}{\bf F}_+({\bf r},t)\times{\bf\nabla}v({\bf r})
 - \frac{1}{2h({\bf r})}{\bf F}_-({\bf r},t)\times{\bf\nabla}
 h({\bf r})\bigr),\nonumber\\
 i\partial_t{\bf F}_-({\bf r},t)
 &=&-v({\bf r})\bigl({\bf\nabla}\times{\bf F}_-({\bf r},t)\label{inh2}\\
 &-&\frac{1}{2v({\bf r})}{\bf F}_-({\bf r},t)\times{\bf\nabla}v({\bf r})
 - \frac{1}{2h({\bf r})}{\bf F}_+({\bf r},t)\times{\bf\nabla}
 h({\bf r})\bigr),\nonumber
\end{eqnarray}
where ${\bf F}_{\pm}({\bf r},t)$ are built with the values of $\epsilon({\bf
r})$ and $\mu({\bf r})$ in the medium, $v({\bf r}) = 1/\sqrt{\epsilon({\bf
r})\mu({\bf r})}$ is the value of the speed of light in the medium, and
$h({\bf r}) = \sqrt{\mu({\bf r})/\epsilon({\bf r})}$ is the ``resistance of
the medium'' (the sole justification for the use of this name is the right
dimensionality of Ohm). The divergence condition (\ref{cmax2}) in an
inhomogeneous medium takes on the form,
\begin{eqnarray}
 {\bf\nabla}\!\cdot\!{\bf F}_+({\bf r},t)
 = \frac{1}{2v({\bf r})}{\bf F}_+({\bf r},t)\!\cdot\!{\bf\nabla}v({\bf r})
 + \frac{1}{2h({\bf r})}{\bf F}_-({\bf r},t)\!\cdot\!{\bf\nabla}h({\bf r}),
 \label{div1}\\
 {\bf\nabla}\!\cdot\!{\bf F}_-({\bf r},t)
 = \frac{1}{2v({\bf r})}{\bf F}_-({\bf r},t)\!\cdot\!{\bf\nabla}v({\bf r})
 + \frac{1}{2h({\bf r})}{\bf F}_+({\bf r},t)\!\cdot\!{\bf\nabla}h({\bf r}).
 \label{div2}
\end{eqnarray}

The quantum-mechanical form of eqs. (\ref{inh1}) and (\ref{inh2}) is
\begin{eqnarray}
 i\hbar\partial_t{\bf F}_+({\bf r},t)
 &=& \sqrt{v({\bf r})}\bigl({\bf s}\!\cdot\!
 \frac{\hbar}{i}{\bf\nabla}\bigr)\sqrt{v({\bf r})}
 {\bf F}_+({\bf r},t)\nonumber\\
 &-& i\hbar\frac{v({\bf r})}{2h({\bf r})}
 \bigl({\bf s}\!\cdot\!{\bf\nabla}h({\bf r})\bigr){\bf F}_-({\bf r},t),
 \label{mfq1}\\
 i\hbar\partial_t{\bf F}_-({\bf r},t)
 &=& \sqrt{v({\bf r})}\bigl({\bf s}\!\cdot\!
 \frac{\hbar}{i}{\bf\nabla}\bigr)\sqrt{v({\bf r})}
 {\bf F}_-({\bf r},t)\nonumber\\
 &+& i\hbar\frac{v({\bf r})}{2h({\bf r})}
 \bigl({\bf s}\!\cdot\!{\bf\nabla}h({\bf r})\bigr){\bf F}_+({\bf r},t).
 \label{mfq2}
\end{eqnarray}

In view of the coupling in the evolution equations (\ref{mfq1}) and
(\ref{mfq2}) between the vectors ${\bf F}_+$ and ${\bf F}_-$, one has to
combine them together to form one wave function ${\cal F}$ with six
components
\begin{eqnarray}
 {\cal F} = \left[\begin{array}{c}{\bf F}_+\\
 {\bf F}_-\end{array}\right].\label{big_wf}
\end{eqnarray}
The wave equation for this function can be written in a compact form
\begin{eqnarray}
 i\hbar\partial_t{\cal F}({\bf r},t)
 &=& \sqrt{v({\bf r})}\rho_3\bigl({\bf s}\!\cdot\!
 \frac{\hbar}{i}{\bf\nabla}\bigr)\sqrt{v({\bf r})}
 {\cal F}({\bf r},t)\nonumber\\
 &+& \hbar\frac{v({\bf r})}{2h({\bf r})}\rho_2
 \bigl({\bf s}\!\cdot\!{\bf\nabla}h\bigr){\cal F}({\bf r},t),
 \label{mf_rho}
\end{eqnarray}
where the spin matrices $s_i$ operate separately on upper and lower
components
\begin{eqnarray}
 s_i{\cal F} = \left[\begin{array}{c}s_i{\bf F}_+\\
 s_i{\bf F}_-\end{array}\right],
\end{eqnarray}
and the three Pauli matrices $\rho_i$ act on ${\cal F}$ as follows
\begin{eqnarray}
 \rho_1{\cal F} = \left[\begin{array}{c}{\bf F}_-\\
 {\bf F}_+\end{array}\right],\;\;
 \rho_2{\cal F} = \left[\begin{array}{c}-i{\bf F}_-\\
 i{\bf F}_+\end{array}\right],\;\;
 \rho_3{\cal F} = \left[\begin{array}{c}{\bf F}_+\\
 -{\bf F}_-\end{array}\right].\label{rho}
\end{eqnarray}
The divergence conditions (\ref{div1}) and (\ref{div2}) can also be written
in this compact form

\begin{eqnarray}
 {\bf\nabla}\!\cdot\!{\cal F}({\bf r},t)
 = \frac{1}{2v({\bf r})}{\cal F}({\bf r},t)\!\cdot\!{\bf\nabla}v({\bf r})
 + \rho_1\frac{1}{2h({\bf r})}{\cal F}({\bf r},t)
 \!\cdot\!{\bf\nabla}h({\bf r}).\label{divf}
\end{eqnarray}
One would not be able to write a linear wave equation for an inhomogeneous
medium in terms of just one three-dimensional complex vector, without
doubling the number of components. Note that the speed of light $v$ may vary
without causing necessarily the mixing of helicities. This happens, for
example, in the gravitational field (cf. \S \ref{curved}). It is only the
space-dependent resistance $h({\bf r})$ that causes mixing.

It is worth stressing that the study of the propagation of photons in an
inhomogeneous medium separates clearly local wave functions from nonlocal
ones. In free space there are many wave functions satisfying the same set of
equations. For example, differentiations of the wave functions or integral
operations of the type (\ref{lp_loc}) do not change the form of these
equations. The essential difference between various wave functions shows up
forcefully in the study of the wave equation in an inhomogeneous medium, or
in curved space (\S \ref{curved}). All previous studies, except
\cite{IBB_94}, were restricted to propagation in {\em free space} and this
very important point was completely missed. The photon wave equations in an
inhomogeneous medium are not very simple but that is due, perhaps, to a
phenomenological character of macroscopic electrodynamics. The propagation
of a photon in a medium is a succession of absorptions and subsequent
emissions of the photon by the charges that form the medium. The number of
photons of a given helicity is, in general, not conserved in these processes
and that accounts for all the complications. The photon wave equations in an
inhomogeneous medium is describing in actual fact a propagation of some
collective excitations of the whole system and not just of free photons.

\subsection{ANALOGY WITH THE DIRAC EQUATION}

The analogy with the relativistic electron theory, mentioned in the
Introduction, becomes the closest when the photon wave equation is compared
with the Dirac equation written in the chiral representation of the Dirac
matrices. In this representation the bispinor is made of two relativistic
spinors
\begin{eqnarray}
 \psi({\bf r},t) = {\phi_A({\bf r},t)\brack \chi^{A'}({\bf r},t)},
 \label{bisp}
\end{eqnarray}
and the Dirac equation may be viewed as two Weyl equations coupled by the
mass term
\begin{eqnarray}
 i\hbar\partial_t \phi({\bf r},t)
 = \;\;c\big(\mbox{\boldmath$\sigma$}\!\cdot\!\frac{\hbar}{i}
 {\bf \nabla}\big)\phi({\bf r},t) + mc^2\chi({\bf r},t),\label{dir1}\\
 i\hbar\partial_t \chi({\bf r},t)
 = -c\big(\mbox{\boldmath$\sigma$}\!\cdot\!\frac{\hbar}{i}
 {\bf \nabla}\big)\chi({\bf r},t) + mc^2\phi({\bf r},t).\label{dir2}
\end{eqnarray}
These equations are analogous to eqs. (\ref{mfq1}) and (\ref{mfq2}) for the
photon wave function. For photons, the role of the mass term is played by
the inhomogeneity of the medium.

\section[Coordinate representation]{Photon wave function in coordinate
representation\label{coordinate}}

Despite a formal similarity between the wave equations for the photon
(\ref{mfq1}) and (\ref{mfq2}) and for the electron (\ref{dir1}) and
(\ref{dir2}), there is an important difference. Photons, unlike the
electrons, do not have antiparticles and this fact influences the form of
solutions of the wave equation and their interpretation.

\subsection{PHOTONS HAVE NO ANTIPARTICLES}

Elementary, plane-wave solutions of relativistic wave equations in free
space are of two types: they have positive or negative frequency,
\begin{eqnarray}
\exp(-i\omega t + {\bf k}\!\cdot\!{\bf r})\;\;\;{\rm or}\;\;\;
\exp(i\omega t - {\bf k}\!\cdot\!{\bf r}).\label{pl_wav}
\end{eqnarray}
According to relativistic quantum mechanics, solutions with positive
frequency correspond to particles and solutions with negative frequency
correspond to antiparticles. Thus, positive and negative frequency parts of
the same solution of the wave equation describe two different physical
entities: particle and antiparticle.

Photons do not have antiparticles, or to put it differently, antiphotons are
identical with photons. Hence, the information carried by the negative
frequency solutions must be the same as the information already contained in
the positive frequency solution. Therefore, one may completely disregard the
negative frequency part as redundant. An alternative method is to keep also
the negative frequency part but to impose an additional condition on the
solutions of the wave equation. This condition states that the operation of
particle-antiparticle conjugation
\begin{eqnarray}
 {\cal F}^c({\bf r},t) = \rho_1{\cal F}^*({\bf r},t),\label{conjugation}
\end{eqnarray}
leaves the function ${\cal F}$ invariant
\begin{eqnarray}
 {\cal F}^c({\bf r},t) = {\cal F}({\bf r},t).\label{conj_inv}
\end{eqnarray}
This condition is compatible with the evolution equation (\ref{mf_rho}) and
it eliminates the unwanted degrees of freedom (cf. \cite{IBB_94}). Eq.
(\ref{conj_inv}) is automatically satisfied if the wave function is
constructed according to the definition (\ref{big_wf}). This follows from
the fact that ${\bf F}_+$ and ${\bf F}_-$ are complex conjugate to each
other. The information carried by the six-component function ${\cal F}$
satisfying the condition (\ref {conj_inv}) is contained in its positive
energy part and is the same as that carried by the initial
Riemann-Silberstein vector ${\bf F}$. It follows from (\ref{conj_inv}) that
the negative frequency part ${\cal F}^{(-)}$ can always be obtained by
complex conjugation and by an interchange of the upper and lower components
of the positive frequency part ${\cal F}^{(+)}$,
\begin{eqnarray}
 {\cal F}^{(-)}({\bf r},t) = \rho_1{\cal F}^{(+)*}({\bf r},t).
 \label{ch_conj}
\end{eqnarray}

In this review, I shall use the symbol $\Psi$ to denote the properly
normalized, {\em positive energy} (positive frequency) part of the function ${\cal
F}$
\begin{eqnarray}
 \Psi({\bf r},t) = {\cal F}^{(+)}({\bf r},t).
\end{eqnarray}
This is the true {\em photon wave function}. Proper normalization of the
photon wave function is essential for its probabilistic interpretation and
is discussed in \S \ref{interpretation}. Note that the function $\Psi$
carries the same amount of information as the original Riemann-Silberstein
vector since $\Psi$ can be constructed from ${\bf F}$ by splitting this
vector into positive and negative frequency parts and then using the first
part as the upper components of $\Psi$ and the complex conjugate of the
second part as the lower components,
\begin{eqnarray}
 \Psi({\bf r},t)
 = {{\bf F}^{(+)}({\bf r},t)\brack {\bf F}^{(-)*}({\bf r},t)}.
 \label{psif}
\end{eqnarray}

The positive frequency part of the solutions of wave equations is a well
known concept also in classical electromagnetic theory where it is called
the analytic signal (\cite{BW_80}, \cite{MW_95}).

\subsection[TRANSFORMATION PROPERTIES]{TRANSFORMATION PROPERTIES OF THE
PHOTON WAVE FUNCTION IN COORDINATE REPRESENTATION}

In free space, the components of the electromagnetic field form a tensor and
that allows one to establish the transformation properties of ${\cal F}$.
Transformation properties of the photon wave function $\Psi$ are the same as
those of ${\cal F}$. Under rotations, the upper and the lower half of $\Psi$
transform as three-dimensional vector fields. Under Lorentz transformations,
the upper and the lower part also transform independently and the
corresponding rules can be inferred directly from classical electrodynamics
(cf., for example, \cite{Jackson_75}). Under the Lorentz transformation
characterized by the velocity ${\bf v}$, the vectors ${\bf F}_{\pm}$ change
as follows
\begin{eqnarray}
 {\bf F}'_{\pm} = \gamma\bigl({\bf F}
 \mp i\frac{{\bf v}\times{\bf F}}{c}\bigr)
 - \frac{\gamma^2}{\gamma + 1}
 \frac{{\bf v}({\bf v}\!\cdot\!{\bf F})}{c^2},
\end{eqnarray}
where $\gamma$ is the standard relativistic factor $\gamma =
\sqrt{1-v^2/c^2}$. One may check that this transformation preserves the
square of these vectors,
\begin{eqnarray}
 ({\bf F}')^2 = ({\bf F})^2.
\end{eqnarray}
This is easily understood if one observes that ${\bf F}^2_{\pm}$ is a
combination of the well-known scalar invariant ${\cal S} = (\epsilon_0{\bf
E}^2 - {\bf B}^2/\mu_0)/2$ and the pseudoscalar invariant ${\cal P} =
\sqrt{\epsilon_0/\mu_0}{\bf E}\!\cdot\!{\bf B}$ of the electromagnetic field
\begin{eqnarray}
 {\bf F}^2_{\pm} = {\cal S} \pm i {\cal P}.
\end{eqnarray}
Thus, rotations and Lorentz transformations act on vectors ${\bf F}_{\pm}$
as elements of the orthogonal group in three dimensions (\cite{Kramers_38}
p. 429) leading to the following transformation properties of the wave
function
\begin{eqnarray}
\Psi'({\bf r}', t') =
\left[\begin{array}{cc}
 C&0\\
 0&C^*\end{array}\right]\Psi({\bf r}, t),\label{c_orth}
\end{eqnarray}
where $C$ is a three-dimensional, complex orthogonal matrix and $C^*$ is its
complex conjugate. The unification of rotations and Lorentz boosts into one
complex orthogonal transformation is even more transparent for infinitesimal
transformations
\begin{eqnarray}
 \Psi' = \Psi + \bigl((i\delta\mbox{\boldmath$\gamma$}
 + \rho_3\delta{\bf v})\!\cdot\!{\bf s}\bigr)\Psi,
\end{eqnarray}
where $\delta\mbox{\boldmath$\gamma$}$ is the vector of an infinitesimal
rotation. Under the space reflection ${\bf r}\to -{\bf r}$, the upper and
lower part of $\Psi$ do not transform independently but are interchanged
because ${\bf D}$ and ${\bf B}$ transform as a vector and a pseudovector,
respectively,
\begin{eqnarray}
 \Psi'(-{\bf r}, t) = \rho_1\Psi({\bf r}, t).
\end{eqnarray}

All these transformation properties can also be simply stated in terms of
second rank spinor fields (cf. \S \ref{spinor}).

\subsection{PHOTON HAMILTONIAN}

The operator appearing on the right hand side of the evolution equation
(\ref {mf_rho}) for the photon wave function is the Hamiltonian operator
$\hat H$ for the photon
\begin{eqnarray}
 {\hat H} = \sqrt{v({\bf r})}\rho_3\bigl({\bf s}\!\cdot\!
 \frac{\hbar}{i}{\bf\nabla}\bigr)\sqrt{v({\bf r})}
 + \hbar\frac{v({\bf r})}{2h({\bf r})}\rho_2
 \bigl({\bf s}\!\cdot\!{\bf\nabla}h({\bf r})\bigr).
 \label{ham_med}
\end{eqnarray}
In free space this expression reduces to
\begin{eqnarray}
 {\hat H}_0 = c\rho_3\bigl({\bf s}\!\cdot\!
 \frac{\hbar}{i}{\bf\nabla}\bigr).\label{ham_free}
\end{eqnarray}
The formulas (\ref {ham_med}) and (\ref {ham_free}) define a Hermitian
operator with continuous spectrum extending from $-\infty$ to $\infty$.
Hermiticity is defined here with respect to the standard (mathematical)
scalar product
\begin{eqnarray}
 (\Psi_1\vert\Psi_2)
 = \int\!d^3r\,\Psi^{\dagger}_1({\bf r})\Psi_2({\bf r}).\label{sc_pr1}
\end{eqnarray}

Wave functions of physical photons are built from positive-energy solutions
of the eigenvalue problem for the Hamiltonian,
\begin{eqnarray}
 {\hat H}\Psi({\bf r}) = E\Psi({\bf r}).\label{eigen_h}
\end{eqnarray}
There is simple relation between positive-energy solutions and
negative-energy solutions. One may obtain all solutions for the negative
energies just by an interchange of upper and lower components since
$\rho_1{\hat H}\rho_1 = -{\hat H}$. Such a simple symmetry of solutions is a
result of photons being identical with antiphotons and it is not found, in
general, for particles whose antiparticles are physically distinct. For
example, the solutions of the wave functions describing electrons in the
Coulomb potential of the proton are quite different from the wave function
of positrons moving in the same potential. In the first case the potential
is attractive (bound states), while in the second case it is repulsive (only
scattering states).

Explicit solutions of the energy eigenvalue problem for photons are easily
obtained in free space but in the presence of a medium this can be done only
in special cases. In this respect, wave mechanics of photons is not much
different from wave mechanics of massive particles, where explicit solutions
can also be found only for special potentials.

\section[Momentum representation]{Photon wave function in momentum
representation \label{momentum}}

The most thorough textbook treatment of quantum mechanics of photons has
been given by \cite{AB_65} who devoted the whole long chapter to this
problem. Their discussion is limited to momentum representation except for a
brief subsection under a characteristic title: "Impossibility of introducing
a photon wave function in the coordinate representation". This impossibility
will be addressed in \S \ref{localizability}.

Wave mechanics of photons in momentum representation can be derived directly
from relativistic quantum kinematics and group representation theory but
here the analysis will be based on the Fourier representation of the photon
wave function in coordinate representation.

In this review I shall use traditionally the wave vector ${\bf k}$ instead
of the photon momentum vector ${\bf p} = \hbar{\bf k}$ as the argument of
the wave function in momentum space. The explicit introduction of Planck's
constant is necessary only when the proper normalization of the wave
function is needed.

\subsection[FOURIER INTEGRAL]{PHOTON WAVE FUNCTION AS A FOURIER INTEGRAL}

The standard procedure for solving the wave equations (\ref{cmax1}) or
(\ref{mf}) is based on Fourier transformation. Since every solution of eqs.
(\ref {cmax1}) and  (\ref {cmax2}) is a solution of the d'Alembert equation
\begin{eqnarray}
 (\frac{1}{c^2}\frac{\partial^2}{\partial t^2}
 - \Delta){\bf F}_{\pm}({\bf r},t) = 0,
\end{eqnarray}
the vectors ${\bf F}_{\pm}$ can be represented as superpositions of plane
waves
\begin{eqnarray}
 {\bf F}_+({\bf r}, t) = \!\sqrt{\hbar c}\!\int\!\!\frac{d^3k}{(2\pi)^3}
 \bigl[{\bf f}_+({\bf k})\,e^{-i\omega t + i{\bf k}\cdot{\bf r}}
 \!+\! {\bf f}^*_-({\bf k})\,e^{i\omega t - i{\bf k}\cdot{\bf r}}\bigr],
 \label{fplus}\\
 {\bf F}_-({\bf r}, t) = \!\sqrt{\hbar c}\!\int\!\!\frac{d^3k}{(2\pi)^3}
 \bigl[{\bf f}_-({\bf k})\,e^{-i\omega t + i{\bf k}\cdot{\bf r}}
 \!+\! {\bf f}^*_+({\bf k})\,e^{i\omega t - i{\bf k}\cdot{\bf r}}\bigr],
 \label{fminus}
\end{eqnarray}
where $\omega = c\vert{\bf k}\vert$ and the factor $\sqrt{\hbar c}$ has been
introduced for future convenience. The remaining integral has the dimension
of $1/{\rm length}^2$, so that the Fourier coefficients ${\bf f}_{\pm}$ have
the dimension of ${\rm length}$. It has already been taken into account in
(\ref{fplus}) and (\ref{fminus}) that the vectors ${\bf F}_+$ and ${\bf
F}_-$ are complex conjugate of each other. In order to fulfill Maxwell's
equations, the two complex vectors ${\bf f}_+({\bf k})$ and ${\bf f}_-({\bf
k})$ must satisfy the set of linear, algebraic equations that result from
(\ref{cmax1}) and (\ref{cmax2}), respectively,
\begin{eqnarray}
 i{\bf k}\times{\bf f}_{\pm}({\bf k})
 &=& \pm\vert{\bf k}\vert{\bf F}_{\pm}({\bf k}),\label{alg1}\\
 {\bf k}\!\cdot\!{\bf f}_{\pm}({\bf k}) &=& 0.\label{alg2}
\end{eqnarray}
Actually, the second equation is superfluous since it follows from the
first. Solutions of these equations are determined up to a complex factor.
Denoting by ${\bf e}({\bf k})$ a normalized solution of the first equation taken with a plus sign
\begin{eqnarray}
 i{\bf k}\times{\bf e}({\bf k})
 &=& \vert{\bf k}\vert{\bf e}({\bf k}),\label{alg3}\\
 {\bf e}^*({\bf k})\!\cdot\!{\bf e}({\bf k}) &=& 1,\label{alg4}
\end{eqnarray}
one can express the vectors ${\bf f}_{\pm}({\bf k})$ in the form
\begin{eqnarray}
 {\bf f}_+({\bf k}) = {\bf e}({\bf k})f({\bf k},1),\;\;\;
 {\bf f}_-({\bf k}) = {\bf e}^*({\bf k})f({\bf k},-1).
\end{eqnarray}
The two complex functions $f({\bf k},\lambda)$, where $\lambda = \pm 1$,
describe the independent degrees of freedom of the free electromagnetic
field. The vector ${\bf e}({\bf k})$ can be decomposed into two real vectors
${\bf l}_i({\bf k})$ that form together with the unit vector ${\bf n}({\bf k}) = {\bf k}/\vert{\bf k}\vert$ an orthonormal set
\begin{eqnarray}
 {\bf e}({\bf k}) = ({\bf l}_1({\bf k})
 + i{\bf l}_2({\bf k}))/\sqrt{2},\;\;\;\;\;
 {\bf l}_i({\bf k})\!\cdot\!{\bf l}_j({\bf k}) = \delta_{ij},\\
 {\bf n}({\bf k})\!\cdot\!{\bf l}_i({\bf k}) = 0,\;\;\;\;\;
 {\bf l}_1({\bf k})\times{\bf l}_2({\bf k}) = {\bf n}({\bf k}).
 \label{ortho}
\end{eqnarray}

The only freedom left in the definition of ${\bf e}({\bf k})$ is its phase.
A multiplication by a phase factor amounts to a rotation of the vectors
${\bf l}_i({\bf k})$ around the vector ${\bf n}({\bf k})$. The same freedom
characterizes the coefficient functions $f({\bf k},\lambda)$. This phase
may, in general, depend on ${\bf k}$ and it plays an important role in the
study of the photon wave function in momentum representation. The final form
of the Fourier representation for vectors ${\bf F}_{\pm}$ is
\begin{eqnarray}
 {\bf F}_+({\bf r}, t)
 = \sqrt{\hbar c}\!\!\int\!\!\frac{d^3k}{(2\pi)^3}{\bf e}({\bf k})
 \bigl[f({\bf k},1)\,e^{-i\omega t + i{\bf k}\cdot{\bf r}}
 + f^*\!({\bf k},-1)\,e^{i\omega t - i{\bf k}\cdot{\bf r}}\bigr]\!,
 \label{fplus1}\\
 {\bf F}_-({\bf r}, t)
 = \sqrt{\hbar c}\!\!\int\!\!\frac{d^3k}{(2\pi)^3}{\bf e}^*\!({\bf k})
 \bigl[f({\bf k},-1)\,e^{-i\omega t + i{\bf k}\cdot{\bf r}}
 + f^*\!({\bf k},1)\,e^{i\omega t - i{\bf k}\cdot{\bf r}}\bigr]\!.
 \label{fminus1}
\end{eqnarray}

\subsection{INTERPRETATION OF FOURIER COEFFICIENTS}

In free space, the energy, momentum, angular momentum, and moment of energy
of the classical electromagnetic field are given by the expressions
(\ref{epm1})--(\ref{epm4}). With the help of the formulas (\ref{fplus1})
of (\ref{fminus1}) they can be expressed in terms of the coefficient functions $f({\bf k},\lambda)$ (cf., for example, \cite{BBBB_75})
\begin{eqnarray}
 E &=& \sum_{\lambda}\int\!\frac{d^3k}{(2\pi)^3\vert{\bf k}\vert}\hbar\omega
 f^*({\bf k},\lambda)f({\bf k},\lambda),\label{en_k}\\
 {\bf P} &=& \sum_{\lambda}\int\!\frac{d^3k}
 {(2\pi)^3\vert{\bf k}\vert}\hbar{\bf k}
 f^*({\bf k},\lambda)f({\bf k},\lambda),\label{mom_k}\\
 {\bf M} &=& \sum_{\lambda}\int\!\frac{d^3k}{(2\pi)^3 \vert{\bf k}\vert}
 f^*({\bf k},\lambda)(\hbar{\bf k}\times\frac{1}{i}{\bf D_k}
 + \lambda\hbar\frac{\bf k}{\vert{\bf k}\vert})f({\bf k},\lambda),
 \label{am_k}\\
 {\bf N} &=& \sum_{\lambda}\int\!\frac{d^3k}{(2\pi)^3 \vert{\bf k}\vert}
 f^*({\bf k},\lambda)i\hbar\omega{\bf D_k}f({\bf k},\lambda),\label{me_k}
 \end{eqnarray}
where
\begin{eqnarray}
 {\bf D_k} = \partial_{\bf k}
 + i\lambda\mbox{\boldmath$\alpha$}({\bf k}),\label{cap_d}
\end{eqnarray}
and
\begin{eqnarray}
 \mbox{\boldmath$\alpha$}({\bf k})
 = {\bf l}_1({\bf k})\cdot\partial_{\bf k}{\bf l}_2({\bf k})
 - {\bf l}_2({\bf k})\cdot\partial_{\bf k}{\bf l}_1({\bf k}).\label{alpha}
\end{eqnarray}
The operation ${\bf D_k}$ is a natural covariant derivative on the light
cone (\cite {Staruszkiewicz_73, BBBB_75, BBBB_87}). Note that this operation
depends through the vector $\mbox{\boldmath$\alpha$}({\bf k})$ on the phase
convention for the polarization vector ${\bf e}({\bf k})$. It is interesting
to note that the vector $\mbox{\boldmath$\alpha$}({\bf k})$ has similar
properties to the electromagnetic vector potential. It changes by a gradient
under a change of the phase, but its curl is uniquely defined. Indeed, it
follows from the definition (\ref{alpha}) of $\mbox{\boldmath$\alpha$}({\bf
k})$ and from the orthonormality conditions (\ref{ortho}) that the vector
$\mbox{\boldmath$\alpha$}({\bf k})$ obeys the equation
\begin{eqnarray}
 \partial_i\alpha_j - \partial_j\alpha_i =
 -\epsilon_{ijk}n_k/{\bf k}^2.\label{berry}
\end{eqnarray}
This equation determines the Berry phase (\cite{BBBB_87}) in the propagation
of photons.

The formulas (\ref{en_k}) and (\ref{mom_k}) indicate that $f({\bf
k},\pm 1)$ describe field amplitudes with energy $\hbar\omega$ and
momentum $\hbar{\bf k}$. The formula (\ref{am_k}) shows that $f({\bf k},1)$
and $f({\bf k},-1)$ describe field amplitudes with positive and negative
helicity since their contribution to the component of angular momentum in
the direction of momentum is equal to $\pm\hbar$, respectively.

The functions $f({\bf k},\lambda)$ have actually a dual interpretation. In
classical theory they yield full information about the electromagnetic
field. In wave mechanics of the photon, these functions are the components
of the photon wave function in momentum representation. In order to
distinguish these two cases, I shall denote the two components of the photon
wave function in momentum representation by a new symbol $\phi({\bf k},
\lambda)$. Wave function must be normalized and the proper normalization of the photon wave function $\phi({\bf k},\lambda)$ is discussed in \S \ref{interpretation}.

The expansion of the photon wave function into plane waves has the following
form
\begin{eqnarray}
 \Psi({\bf r}, t) = \sqrt{\hbar c}\!\int\!
 \frac{d^3k}{(2\pi)^3}{{\bf e}({\bf k},1)\phi({\bf k},1)\brack
 {\bf e}({\bf k},-1)\phi({\bf k},-1)}
 e^{-i\omega t + i{\bf k}\cdot{\bf r}},\label{gen_psi}
\end{eqnarray}
where
\begin{eqnarray}
 {\bf e}({\bf k},1) = {\bf e}({\bf k}),
 \;\;\;{\bf e}({\bf k},-1) = {\bf e}^*({\bf k}).\label{e1e2}
\end{eqnarray}
The integral (\ref{gen_psi}) defines a certain (continuous) superposition
of the wave functions $\phi({\bf k},\lambda)$ with
different values of the wave vector. If the photon wave function in momentum
representation is accepted as a legitimate concept, then the superpositions
of such functions must also be accepted. Those who are not sure about the
meaning of a continuous superpositions, in the form of an integral, may
restrict the electromagnetic field to a box and replace the integral by a
discrete sum.

\subsection[TRANSFORMATION PROPERTIES IN MOMENTUM SPACE]{TRANSFORMATION
PROPERTIES OF THE PHOTON WAVE FUNCTION IN MOMENTUM REPRESENTATION}

Transformation rules for the wave function in momentum representation may be
derived from the transformation properties of the Riemann-Silber\-stein
vector. They are the same in the classical theory of the electromagnetic
field and in the quantum theory of photons. Under space and time
translations, functions $\phi({\bf k},\lambda)$ are multiplied by the phase
factors
\begin{eqnarray}
 \phi'({\bf k},\lambda)
 = \exp(-i\omega t_0 + i{\bf k}\!\cdot\!{\bf r}_0)\phi({\bf k},\lambda),
\end{eqnarray}
where $({\bf r}_0, t_0)$ is the four-vector of translation. From expressions (\ref{en_k}) and (\ref{mom_k}) one may deduce that under rotations and
Lorentz transformations, the functions $\phi({\bf k},\lambda)$ are also
multiplied by phase factors
\begin{eqnarray}
 \phi'({\bf k}',\lambda) = \exp(-i\lambda\Theta({\bf k},\Lambda))
 \phi({\bf k},\lambda),\label{ftransf}
\end{eqnarray}
where the phase function $\Theta({\bf k},\Lambda)$ depends on the Poincar\'e
transformation $\Lambda$. This transformation property can easily be derived
from the transformation law of the energy-momentum four-vector for the
electromagnetic field. The right hand side in the formulas (\ref{en_k}) and
(\ref{mom_k}) has three factors: the integration volume
$d^3k/(2\pi)^3\vert{\bf k}\vert$, the four-vector $({\bf k},\omega)$, and
the moduli squared of $\phi({\bf k},\lambda)$. The integration volume is an
invariant (cf., for example, \cite{Weinberg_95}, p. 67) and therefore
$\vert\phi({\bf k},\lambda)\vert^2$ must also be invariant. An explicit form
of the phase $\Theta({\bf k},\Lambda)$ corresponding to a given Poincar\'e
transformation $\Lambda$ can be given (cf., for example, \cite{Amrein_69})
but it is not very illuminating.

\section{Probabilistic interpretation\label{interpretation}}

Probabilistic interpretation of wave mechanics requires, first of all, a
definition of the scalar product $\langle\Psi_1\vert\Psi_2\rangle$ that is
to be used in the calculation of transition probabilities. The modulus
squared of the scalar product of two normalized wave functions
$\vert\langle\Psi_1\vert\Psi_2\rangle\vert^2$ determines the probability of finding
a photon in the state $\Psi_1$ when the photon is in the state $\Psi_2$. The
probability, of course, must be a pure number and --- as a true observable --- must be
invariant under all Poincar\'e transformations. The most obvious definition
of the scalar product (\ref{sc_pr1}) can not be used because it is neither
Poincar\'e invariant nor dimensionally correct. There is essentially only
one candidate for the correct scalar product. Its heuristic derivation is
the easiest in momentum representation.

\subsection[SCALAR PRODUCT]{SCALAR PRODUCT}

According to quantum mechanics each photon with momentum $\hbar{\bf k}$
carries energy $\hbar\omega$. Thus, the total number of photons $N$ present
in the electromagnetic field is obtained by dividing the integrand in the
formula (\ref{en_k}) by $\hbar\omega$,
\begin{eqnarray}
 N = \sum_{\lambda}\int\!\frac{d^3k}{(2\pi)^3\vert{\bf k}\vert}
 \vert f({\bf k},\lambda)\vert^2.\label{num_ph}
\end{eqnarray}
A photon wave function describes just one photon. The normalized wave
function must, therefore, satisfy the condition $N = 1$. Normalized photon
wave functions in momentum representation satisfy the normalization
condition
\begin{eqnarray}
 \sum_{\lambda}\int\!\frac{d^3k}{(2\pi)^3\vert{\bf k}\vert}
 \vert\phi({\bf k},\lambda)\vert^2 = 1.
\end{eqnarray}

The form of the scalar product can be deduced from the expression for the
norm and it reads
\begin{eqnarray}
 \langle\Psi_1\vert\Psi_2\rangle
 = \sum_{\lambda}\!\int\!\frac{d^3k}{(2\pi)^3\vert{\bf k}\vert}
 \phi^*_1({\bf k},\lambda)\phi_2({\bf k},\lambda).\label{sc_pr_k}
\end{eqnarray}
This scalar product can be also expressed in terms of the photon wave
functions in coordinate representation by inverting the Fourier
transformation in eq. (\ref{gen_psi})
\begin{eqnarray}
 \phi({\bf k},\lambda) = \frac{1}{\sqrt{\hbar c}}
 {\bf e}^*({\bf k},\lambda)\!\cdot\!
 \int\!d^3r\exp(-i{\bf k}\!\cdot\!{\bf r})\Psi({\bf r},t),
 \label{inv_f}
\end{eqnarray}
where the scalar product with ${\bf e}^*({\bf k},\lambda)$ is evaluated for
upper and lower components separately and for each $\lambda$ only one of
them does not vanish. Upon substituting this expression into
(\ref{sc_pr_k}), interchanging the order of integrations, and using the
following properties of the vectors ${\bf e}({\bf k},\lambda)$
\begin{eqnarray}
 {\bf e}^*({\bf k},\lambda)\!\cdot\!{\bf e}({\bf k},\lambda')
 &=& \delta_{\lambda,\lambda'},\label{elel}\\
 \sum_{\lambda}e^*_i({\bf k},\lambda)e_j({\bf k},\lambda)
 &=& \delta_{ij} - n_i({\bf k})n_j({\bf k}),
\end{eqnarray}
one obtains the following expression for the scalar product in coordinate
representation
\begin{eqnarray}
 \langle\Psi_1\vert\Psi_2\rangle
 = \frac{1}{2\pi^2\hbar c}\int\!d^3r\int\!d^3r'\Psi^{\dagger}_1
 ({\bf r})\frac{1}{\vert{\bf r - r'}\vert^2}\Psi_2({\bf r'}).
 \label{sc_pr_r}
\end{eqnarray}
The norm associated with this scalar product is
\begin{eqnarray}
 N = \Vert\Psi\Vert^2
 = \frac{1}{2\pi^2\hbar c}\int\!d^3r\int\!d^3r'
 \Psi^{\dagger}({\bf r})\frac{1}{\vert{\bf r - r'}\vert^2}\Psi({\bf r'}).
 \label{norm_r}
\end{eqnarray}

The scalar product (\ref{sc_pr_r}) and the associated norm (\ref{norm_r})
for photon wave functions have been arrived at by numerous authors starting
from various premises. \cite{Gross_64} has proven that this scalar product
and this norm are invariant not only under Poincar\'e transformations but
also under conformal transformations. \cite{Zeldovich_65} derived the
formula (\ref{norm_r}) for the number of photons in terms of the
electromagnetic field vectors. Recently, the norm (\ref{norm_r}) has been found very useful in the formulation of wavelet electrodynamics (\cite{Kaiser_92}). The same expression (\ref{sc_pr_k}) for the
scalar product can also be derived by considering quantum-mechanical
expectation values. That approach has been used by \cite{Good_57} and is presented below.

\subsection[EXPECTATION VALUES]{EXPECTATION VALUES OF PHYSICAL QUANTITIES}

In wave mechanics of photons, the normalized photon wave function $\phi({\bf
k},\lambda)$ replaces the classical field amplitudes $f({\bf k},\lambda)$.
The classical expressions for the energy, momentum, angular momentum, and
moment of energy become the formulas for the quantum-mechanical expectation
values
\begin{eqnarray}
 \langle E\rangle = \langle\Psi\vert{\hat H}\vert\Psi\rangle,\;\;
 \langle{\bf P}\rangle = \langle\Psi\vert{\hat{\bf P}}\vert\Psi\rangle,\\
 \langle{\bf M}\rangle = \langle\Psi\vert{\hat{\bf J}}\vert\Psi\rangle,\;\;
 \langle{\bf N}\rangle = \langle\Psi\vert{\hat{\bf K}}\vert\Psi\rangle.
 \label{expect}
\end{eqnarray}
These equations compared with the formulas (\ref{en_k})--(\ref{me_k}) enable
one to identify the operators ${\hat H}$, ${\hat{\bf P}}$, ${\hat{\bf J}}$,
and ${\hat{\bf K}}$ in momentum representation as
\begin{eqnarray}
 {\hat H} &=& \hbar\omega,\label{epm1_k}\\
 {\hat{\bf P}} &=& \hbar{\bf k},\label{epm2_k}\\
 {\hat{\bf J}} &=& {\bf k}\times\frac{\hbar}{i}{\bf D_k}
 + \lambda\hbar\frac{\bf k}{\vert{\bf k}\vert},\label{epm3_k}\\
 {\hat{\bf K}} &=& \hbar\omega\frac{\hbar}{i}{\bf D_k}.\label{epm4_k}
\end{eqnarray}

The operators ${\hat H}$, ${\hat{\bf P}}$, ${\hat{\bf J}}$, and
${\hat{\bf K}}$ are Hermitian with respect to the scalar product given by the formula (\ref{sc_pr_k}).

The formulas (\ref{epm1_k})--(\ref{epm4_k}) are fully consistent with the
interpretation of $\phi({\bf k},\lambda)$ as the probability amplitude in
momentum representation. The probability density to find the photon with the
momentum $\hbar{\bf k}$ and the helicity $\lambda$ is
\begin{eqnarray}
 {\rm Probability\; density}
 = \frac{\vert{\phi({\bf k},\lambda })\vert^2}{(2\pi)^3\vert{\bf k}\vert}.
\end{eqnarray}

In order to express the quantum-mechanical expectation values
(\ref{epm1_k})--(\ref{epm4_k}) in coordinate representation one must
identify the proper form of the scalar product for the photon wave function
$\Psi$. This identification has already been made by \cite{Good_57} who
compared the classical formula for the energy of the electromagnetic field
with the quantum-mechanical expression involving the Hamiltonian
\begin{eqnarray}
 \langle E\rangle = \langle\Psi\vert{\hat H}\vert\Psi\rangle,
\end{eqnarray}
and came to the conclusion that the scalar product for the photon wave
function has to be modified as follows
\begin{eqnarray}
\langle\Psi_1\vert\Psi_2\rangle
 = \int\!d^3r\,\Psi^{\dagger}_1\frac{1}{\hat H}\Psi_2.\label{sc_pr_h}
\end{eqnarray}
It is assumed here that the wave functions are built from positive energy
states only and that guarantees the positive definiteness of the norm
\begin{eqnarray}
\Vert\Psi\Vert^2
 = \int\!d^3r\,\Psi^{\dagger}\frac{1}{\hat H}\Psi,\label{norm_h}
\end{eqnarray}
associated with that scalar product. This form of the scalar product leads
to the following expectation values of the energy, momentum, angular
momentum, and moment of energy operators
\begin{eqnarray}
 \langle E\rangle
 &=&  \int\!d^3r\,\Psi^{\dagger}{\hat H}^{-1}
 \Big({\bf s}\!\cdot\!\frac{\hbar}{i}{\bf\nabla}\Big)\Psi,
 \label{en_ev}\\
 \langle{\bf P}\rangle
 &=& \int\!d^3r\,\Psi^{\dagger}{\hat H}^{-1}\frac{\hbar}{i}{\bf\nabla}\Psi,
 \label{mom_ev}\\
 \langle{\bf M}\rangle
 &=& \int\!d^3r\,\Psi^{\dagger}{\hat H}^{-1}\big({\bf r}
 \times\frac{\hbar}{i}{\bf\nabla} + \hbar{\bf s}\big)\Psi,\label{amom_ev}\\
 \langle{\bf N}\rangle
 &=& \int\!d^3r\,\Psi^{\dagger}{\hat H}^{-1}
 {\hat H}{\bf r}\Psi.\label{me_ev}
 \end{eqnarray}
Thus, the operators ${\hat H}$, ${\hat{\bf P}}$, ${\hat{\bf J}}$, and
${\hat{\bf K}}$ in coordinate representation have the form
\begin{eqnarray}
 {\hat H}
 &=& c\Big({\bf s}\!\cdot\!\frac{\hbar}{i}{\bf\nabla}\Big),\label{epm1_r}\\
 {\hat{\bf P}} &=& \frac{\hbar}{i}{\bf\nabla},\label{epm2_r}\\
 \hat{\bf J}
 &=& {\bf r}\times\frac{\hbar}{i}{\bf\nabla} + \hbar{\bf s},\label{epm3_r}\\
 {\hat{\bf K}} &=& {\hat H}{\bf r}.\label{epm4_r}
\end{eqnarray}
All these operators preserve the divergence condition (\ref{cmax2}) and they
are Hermitian with respect to the scalar product (\ref{sc_pr_h}). It is also
reassuring to note that the quantum-mechanical operators of momentum and
angular momentum in coordinate representation have the same form as in
standard quantum mechanics. This can be taken as another indication that
$\Psi({\bf r}, t)$ is a legitimate and useful object.

One may prove directly (without using the Fourier expansions) with the help
of the following identities
\begin {eqnarray}
 \hat H \rho_3 {\bf s}/c \Psi = \frac{\hbar}{i}{\bf\nabla} \Psi,\;\;\;
 \hat H \rho_3 \hat{\bf r}\times{\bf s}/c \Psi
 = (\hat{\bf r}\times\frac{\hbar}{i}{\bf\nabla}
 + \hbar{\bf s}) \Psi,\label{rel}
\end {eqnarray}
and with the use of eq. (\ref{aux_s}) that the expectation values
(\ref{en_ev})--(\ref{amom_ev}) reduce to the classical expressions
(\ref{epm1})--(\ref{epm4}) when the wave function is replaced by the
classical electromagnetic field.

The scalar product (\ref{sc_pr_h}) in coordinate representation has been
obtained from the scalar product (\ref{sc_pr_k}) in momentum representation.
However, the scalar product that contains the division by the Hamiltonian
can be derived on more general grounds and its definition does not depend on
the choice of representation. It has been shown (\cite{Segal_63}, \cite{AM_75}) that such a scalar product is a general feature of geometric quantization in field theory.

Even though the number of photons is given by a double integral, so that
there is no local expression for the photon probability density in
coordinate space, the expression for the energy has the form of a single
integral over $\vert\Psi({\bf r})\vert^2$. Therefore, one may introduce a
tentative notion of the "average photon energy in a region of space" and try
to associate a probabilistic interpretation of the photon wave function with
this quantity (\cite{IBB_94}, \cite{Sipe_95}). More precisely, the quantity
$p_E(\Omega)$,
\begin{eqnarray}
 p_E(\Omega)
 = \frac{\int_{\Omega}\!d^3r\Psi^{\dagger}({\bf r})\Psi({\bf r})}
 {\langle E\rangle},
\end{eqnarray}
may be interpreted as the probability to find the energy of the photon
localized in the region $\Omega$. In other words, $p_E(\Omega)$ is the
fraction of the average total energy of the photon associated with the
region $\Omega$. The probability density $\rho_E({\bf r},t)$ to find the
energy of the photon at the point ${\bf r}$,
\begin{eqnarray}
 \rho_E({\bf r},t)
 = \frac{\Psi^{\dagger}({\bf r}, t)\Psi({\bf r}, t)}{\langle E\rangle},
\end{eqnarray}
is properly normalized to one and it also satisfies the continuity equation
\begin{eqnarray}
 \partial_t\rho_E({\bf r},t) + {\bf\nabla}\!\cdot\!{\bf j}_E({\bf r},t) = 0,
\end{eqnarray}
with the normalized average energy flux
\begin{eqnarray}
 {\bf j}_E({\bf r},t)
 = \frac{\Psi^{\dagger}({\bf r}, t)\rho_3{\bf s}\Psi({\bf r}, t)}
 {\langle E\rangle},
\end{eqnarray}
as the probability current. The direct connection between the wave function
$\Psi({\bf r},t)$ and the average energy density justifies the name "the
energy wave function" used by \cite{MW_95}. It is understandable that the
localization of photons is associated with their energy because photons do
not carry other attributes like charge, fermion number, or rest mass. It is
worth noting that for gravitons not only the probability but even the energy
can not be localized (cf., for example, \cite{WW_80}). The probabilistic
interpretation of the energy wave function $\Psi$ is still subject to all
the limitations arising from the lack of the photon position operator, as
discussed in \S \ref{localizability}. In particular, there are no projection
operators whose expectation values would give the probabilities
$p_E(\Omega)$.

The transition amplitudes, the operators representing important physical
quantities, and the expectation values can be expressed with equal ease in
momentum representation and in coordinate representation. For photons moving
in empty space both representations are completely equivalent and give the
same results. The only relevant issue is whether a particular superposition
of wave functions in momentum representation is useful for the description
of quantum states of the photon. The distinguished and unique feature of the
superposition given by the Fourier integrals (\ref{gen_psi}) is that they
represent {\em local} fields. They have local transformation properties
(\ref{c_orth}) and they satisfy local boundary conditions. Therefore, for
photons moving in an inhomogeneous or bounded medium, it is the coordinate
representation that is preferred because only in this representation one may
easily take into account the properties of the medium (cf. \S
\ref{eigenvalue}).

\subsection{CONNECTION WITH LANDAU-PEIERLS WAVE FUNCTION}

One may easily convert the scalar product (\ref{sc_pr_r}) into a standard
form containing a single integration with the use of the following identity
\begin{eqnarray}
 \frac{1}{16\pi}\int\!d^3r\frac{1}{\vert{\bf r - r'}\vert^{5/2}}
 \frac{1}{\vert{\bf r - r''}\vert^{5/2}}
 = \frac{1}{\vert{\bf r' - r''}\vert^2}.
\end{eqnarray}
This enables one to convert the double integral (\ref{sc_pr_r}) into a
single integral
\begin{eqnarray}
 \langle\Psi_1\vert\Psi_2\rangle
 = \int\!d^3r\Phi^{\dagger}_1({\bf r})\Phi_2({\bf r}).\label{sc_lp}
\end{eqnarray}
The new functions $\Phi$ are the Landau-Peierls wave functions and they are
related to the photon wave functions $\Psi$ through the formula
\begin{eqnarray}
 \Phi({\bf r}) = \frac{\pi}{\sqrt{\hbar c}}\int\!d^3r'
 \frac{1}{(2\pi\vert{\bf r - r'}\vert)^{5/2}}\Psi({\bf r'}).\label{lp_wf}
\end{eqnarray}
The form of the scalar product for the Landau-Peierls wave functions is
simple but one must pay for this simplicity with the nonlocality of the wave
functions. There is also a simple mathematical argument that shows
shortcomings of the Landau-Peierls wave function. While for every integrable
wave function $\Psi$ the transformation (\ref{lp_wf}) defines the
Landau-Peierls wave function $\Phi$, the inverse transformation is singular
since it contains a nonintegrable kernel $\vert{\bf r - r'}\vert^{7/2}$
(\cite{Amrein_69}, \cite{MW_95}). This leads to a paradox that for many
"reasonable" functions $\Phi$ (for example, for every function that becomes
zero abruptly at the boundary) the energy density is infinite. Thus, it is
much more natural to treat $\Psi$ as the primary and $\Phi$ as the derived
object.

\section[Eigenvalue problems]{Eigenvalue problems for the photon wave
function\label{eigenvalue}}

In wave mechanics of photons as in wave mechanics of massive particles, one
may study eigenvalues and eigenfunctions of various interesting observables.
The most important observables, of course, are the momentum, angular
momentum, energy, and moment of energy --- the generators of the Poincar\'e
group. The eigenfunctions of these observables will be given in coordinate
representation to underscore the validity and usefulness of the photon wave
function in this representation.

\subsection[MOMENTUM AND ANGULAR MOMENTUM]{EIGENVALUE PROBLEMS FOR MOMENTUM
AND ANGULAR MOMENTUM}

The eigenvalue problems for the components of the photon momentum operator
have the standard quantum-mechanical form
\begin{eqnarray}
\hat P_i\;\Psi({\bf r}) = \hbar k_i\;\Psi({\bf r}),
\end{eqnarray}
and its solutions depend on ${\bf r}$ through the exponential functions
$\exp(i{\bf k}\!\cdot\!{\bf r})$.

The eigenvalue problem for the photon angular momentum also has the standard
quantum-mechanical form. It contains, as usual, the eigenvalue problem for
the $z$-component of the total angular momentum
\begin{eqnarray}
\hat J_z\,\Psi({\bf r}) = \hbar M\Psi({\bf r}),\label{ezam}
\end{eqnarray}
and the eigenvalue problem for the square of the total angular momentum
\begin{eqnarray}
\hat{\bf J}^2\,\Psi({\bf r}) = \hbar^2 J(J+1)\,\Psi({\bf r}).\label{etam}
\end{eqnarray}
The solutions of eqs. (\ref{ezam}) and (\ref{etam}) are well known vector
spherical harmonics (cf., for example, \cite{Messiah_61}). The direct
connection between the quantum-mechanical eigenvalue problems and multipole
expansion in classical electromagnetism has been explored systematically for
the first time by \cite{Moliere_49}.

\subsection[MOMENT OF ENERGY]{EIGENVALUE PROBLEM FOR THE MOMENT OF ENERGY}

The solution of the eigenvalue problem for the moment of energy shows the
versatility of the calculational methods based on the coordinate
representation and sheds some light on the problem of the localizability of
the photon that is discussed in \S \ref{localizability}. Of course, the same
result can be obtained by Fourier transforming the solution of the eigenvalue
problem obtained in momentum representation.

The three components of the moment of energy, like the components of angular
momentum, do not commute among themselves. Therefore, the eigenvalue problem
can be posed only for one component at a time. Choosing, for definiteness,
the $z$-component, one obtains the following eigenvalue equation
\begin{eqnarray}
 -i\rho_3\big({\bf s}\!\cdot\!{\bf\nabla}\big)z\,\Psi = \kappa\,\Psi.
 \label{kappa}
\end{eqnarray}
The solution of this equation becomes unique when one chooses two additional
eigenvalue equations to be solved concurrently. For example, eq.
(\ref{kappa}) can be solved together with the eigenvalue problems for the
$x$ and $y$ components of momentum since the three operators ${\hat P}_x$,
${\hat P}_y$, and ${\hat K}_z$ commute. The solutions for the upper and
lower components of the wave function differ only in the sign of $\kappa$
and one can solve them independently. When the wave function in the form
$\Psi=\exp(ik_xx + ik_yy)(\psi_x(z),\psi_y(z),\psi_z(z))$ is substituted
into (\ref{kappa}), one obtains a set of ordinary differential equations
\begin{eqnarray}
 ik_y z\psi_z - (z\psi_y)' = \kappa\,\psi_x,\\
 (z\psi_x)' - ik_x z\psi_z = \kappa\,\psi_y,\\
 ik_x z\psi_y - ik_y z\psi_x = \kappa\,\psi_z,
\end{eqnarray}
where the prime denotes the differentiation with respect to $z$. These
equations are solved by the following substitution ($k^2_{\perp} = k^2_x +
k^2_y$)
\begin{eqnarray}
 \psi_x = \frac{i}{k^2_{\perp}z}(k_y\kappa + k_x\frac{d}{dz})\psi_z,
 \label{psi_x}\\
 \psi_y = \frac{i}{k^2_{\perp}z}(-k_x\kappa + k_y\frac{d}{dz})\psi_z,
 \label{psi_y}
\end{eqnarray}
which results in a Bessel-type equation for $\psi_z$,
\begin{eqnarray}
 z^2\psi''_z + z\psi'_z + (\kappa^2 - k^2_{\perp}z^2)\psi_z = 0.
\end{eqnarray}
The physically acceptable solution of this equation is given by the
Macdonald function of the imaginary index
\begin{eqnarray}
 \psi_z(z) = K_{i\kappa}(k_{\perp}z) = \int_0^{\infty}\!dt\,
 e^{-k_{\perp}z\cosh t}\cos(\kappa t).\label{macdon}
\end{eqnarray}
The other solution grows exponentially when $z\to\infty$ and must be
rejected. The physical solution falls off exponentially for large
$\vert{z}\vert$ and represents a photon state that is localized as much as
possible in the $z$-direction. The remaining two components of the
eigenfunction are obtained from eqs. (\ref{psi_x}) and (\ref{psi_y}). The
photon wave functions that describe eigenstates of ${\hat K}_z$ are not
normalizable, because the spectrum of the eigenvalues is continuous:
$\kappa$ can be any real number.

\subsection[PROPAGATION IN OPTICAL FIBER]{PHOTON PROPAGATION ALONG AN OPTICAL FIBER AS A QUANTUM MECHANICAL BOUND STATE PROBLEM}
The eigenvalue problem for the photon energy operator in the absence of a
medium is solved by Fourier transformation as described in \S
\ref{momentum}. In the presence of a medium, one can search for eigenstates
of the photon Hamiltonian closely following the path traveled in
nonrelativistic wave mechanics of massive particles. This procedure usually
involves selecting a set of operators commuting with the Hamiltonian and
then solving the appropriate set of eigenvalue equations. The photon
propagation along an infinite cylindrical optical fiber (cf., for example,
\cite{IBB_94}) is a good illustration of this approach. In order to take
care of the boundary conditions at the surface of the fiber, one must work
in the coordinate representation.

Consider an infinite, cylindrical optical fiber of diameter $a$
characterized by a dielectric permittivity $\epsilon$. The symmetry of the
problem suggests the inclusion in the set of commuting operators, in
addition to the Hamiltonian, the projections of the momentum operator and
the total angular momentum on the direction of the fiber axis. In
cylindrical coordinates the eigenvalue equations for the $z$-components of
momentum and angular momentum and the Hamiltonian have the form
\begin{eqnarray}
-i\partial_z\,\Psi &=& k_z\Psi,\label{eigen1}\\
 \big(-i\partial_{\varphi} + ({\bf s}\!\cdot\!{\bf e}_z)\big)\,\Psi
 &=& M\Psi,\label{eigen2}\\
 -i\rho_3\big({\bf s}\!\cdot\!({\bf e}_{\rho}\partial_{\rho}
 + \frac{1}{\rho}{\bf e}_{\varphi}\partial_{\varphi}
 + {\bf e}_{z}\partial_{z})\big)\,\Psi
 &=& \frac{\omega}{v}\,\Psi,\label{eigen3}
\end{eqnarray}
where ${\bf e}_{\rho}$, ${\bf e}_{\varphi}$, and ${\bf e}_{z}$ are the unit
vectors along the coordinate lines and $v$ equals to $c$ outside the fiber.
Due to the symmetry of the problem, there is no coupling between the upper
and lower components of $\Psi$ and the solution of these eigenvalue
equations can be sought in the form of a three dimensional vector,
\begin{eqnarray}
 \Psi = {\bf e}_{\rho}\psi_{\rho} + {\bf e}_{\varphi}\psi_{\varphi}
 + {\bf e}_{z}\psi_z.
\end{eqnarray}
In order to separate the variables and obtain a set of ordinary differential
equations, one needs the following differential and algebraic relations
\begin{eqnarray}
 \partial_{\varphi}{\bf e}_{\rho} = {\bf e}_{\varphi},\;\;
 \partial_{\varphi}{\bf e}_{\varphi} = -{\bf e}_{\rho},\label{diff}\\
 ({\bf s}\!\cdot\!{\bf e}_{\rho}){\bf e}_{\varphi} = i{\bf e}_z,\;\;
 ({\bf s}\!\cdot\!{\bf e}_z){\bf e}_{\rho} = i{\bf e}_{\varphi},\;\;
 ({\bf s}\!\cdot\!{\bf e}_{\varphi}){\bf e}_z = i{\bf e}_{\rho}.
 \label{algeb}
\end{eqnarray}
All unlisted terms of the type (\ref{diff}) and (\ref{algeb}) vanish. The
dependence on $\varphi$ and $z$ of all three components $\psi_{\rho}$,
$\psi_{\varphi}$, and $\psi_z$ of the photon wave function can be separated
out on the basis of eqs. (\ref{eigen1}) and (\ref{eigen2}),
\begin{eqnarray}
\psi = \exp(ik_z z)\exp(iM\varphi) f(\rho).
\end{eqnarray}
The three $\rho$-dependent components of the wave function satisfy the
equations
\begin{eqnarray}
-\frac{M}{\rho}f_z + k_z f_{\varphi} &=&
\frac{\omega}{v}(if_{\rho}),\label{cyl1}\\
-\partial_{\rho} f_z + k_z (if_{\rho}) &=&
\frac{\omega}{v}f_{\varphi},\label{cyl2}\\
\frac{1}{\rho}\partial_{\rho}\rho f_{\varphi} -
\frac{M}{\rho}(if_{\rho}) &=& \frac{\omega}{v}f_z.\label{cyl3}
\end{eqnarray}
These equations lead to a Bessel equation for $f_z$
\begin{eqnarray}
\left[\partial_{\rho}^2 + \frac{1}{\rho}\partial_{\rho} -
\frac{m^2}{\rho^2} + k_{\perp}^2\right]f_z = 0,\label{bessel}
\end{eqnarray}
where $k_{\perp}^2 = \omega^2/v^2 - k_z^2$. The remaining two functions $f$
can be determined in terms of $f_z$,
\begin{eqnarray}
 f_{\rho}
 &=& ik^{-2}_{\perp}\big(\frac{\omega M}{v\rho}
 + k_z\partial_{\rho}\big)f_z,\\
 f_{\varphi}
 &=& k^{-2}_{\perp}\big(\frac{M k_z}{\rho}
 + \frac{\omega}{v}\partial_{\rho}\big)f_z.
\end{eqnarray}
The photon wave function obeys the Bessel equation inside the fiber with one
value of $k_{\perp}$ and with a different values of $k_{\perp}$ in the surrounding free space. The behavior of the solution of eq. (\ref{bessel})
depends on whether $k_{\perp}$ is real or imaginary. A general solution of
this equation is either (for real $k_{\perp}$) a linear combination of Bessel functions of the first kind $J_M(\rho)$ and the second kind $Y_M(\rho)$ or
(for imaginary $k_{\perp}$) a linear combination of modified Bessel functions $I_M(\rho)$ and $K_M(\rho)$. In full analogy with the problem of a potential well in quantum mechanics, one can search for bound states in the transverse direction by matching a regular oscillatory solution inside (i.e. the $J_M(\rho)$ function) with an exponentially damped solution outside the fiber (i.e. the $K_M(\rho)$ function). The matching conditions, well known from classical electromagnetic theory, are the continuity conditions for the $E_z$ and $H_z$ field components at the surface of the fiber, when $\rho = a$. Bound states occur because the speed of light is greater in the vacuum than inside the fiber. Therefore, it may happen that $k_{\perp}$ is real inside and imaginary outside the fiber. Since there are two matching conditions and only one ratio of the amplitudes inside and outside the fiber, both conditions can be satisfied only for a set of discrete eigenvalues of the photon energy $\hbar\omega$. It is worth noting that in order to have an imaginary $k_{\perp}$ one must have a nonvanishing $k_z$. Thus, a photon may be bound in the plane perpendicular to the fiber, but it is always moving freely along the fiber, as in the quantum-mechanical description of a charged particle moving in a homogeneous magnetic field. This analysis gives an interpretation of electromagnetic evanescent waves as quantum bound states. Of course, true bound states of photons, that are described by a photon wave function decaying exponentially in {\em all} directions, are not possible.

\section[RELATIVISTIC INVARIANCE]{Relativistic invariance of photon wave
mechanics\label{invariance}}

In a relativistically invariant quantum theory, the Poincar\'e
transformations are represented by unitary operators. The ten Hermitian
generators of these transformations must satisfy the commutation relations
characteristic of the Poincar\'e group. The ten generators of the Poincar\'e
group are identified with the operators ${\hat H}$, ${\hat{\bf P}}$,
${\hat{\bf J}}$, and ${\hat{\bf K}}$. They generate infinitesimal time
translation, space translations, rotations, and boosts (special Lorentz
transformations), respectively. The structure of the Poincar\'e group leads
to the following commutation relations obeyed by these generators (cf., for
example, \cite{BW_48}, \cite{BBBB_75}, \cite{IZ_80}, \cite{Weinberg_95}, p.
61)
\begin{eqnarray}
 \left[{\hat J}_i,{\hat P}_j\right] &=& i\hbar \epsilon_{ijk}{\hat P}_k,
 \label{cmp}\\
 \left[{\hat J}_i,{\hat J}_j\right] &=& i\hbar \epsilon_{ijk}{\hat J}_k,
 \label{cmm}\\
 \left[{\hat J}_i,{\hat K}_j\right] &=& i\hbar \epsilon_{ijk}{\hat K}_k,
 \label{cmn}\\
 \left[{\hat K}_i,{\hat P}_j\right] &=& i\hbar  c^{-2}\delta_{ij}{\hat H},
 \label{cnp}\\
 \left[{\hat K}_i,{\hat H}\right] &=& i\hbar {\hat P}_i,\label{cne}\\
 \left[{\hat K}_i,{\hat K}_j\right]
 &=& -ic^{-2}\hbar\epsilon_{ijk}{\hat J}_k.\label{cnn}
\end{eqnarray}
All the remaining commutators vanish. One may check by a direct calculation
that the operators ${\hat H}$, ${\hat{\bf P}}$, ${\hat{\bf J}}$, and
${\hat{\bf K}}$, given in momentum representation by the formulas
(\ref{epm1_k})--(\ref{epm4_k}) and in coordinate representation by the
formulas (\ref{epm1_r})--(\ref{epm4_r}), obey the commutation relation for
the generators of the Poincar\'e group. In the proof of the commutation
relations in momentum representation one needs the condition (\ref{berry}).
Since all generators of the Poincar\'e group are represented by operators
that are Hermitian with respect to the scalar product (\ref{sc_pr_k}) or
(\ref{sc_pr_r}), the Poincar\'e transformations are represented by unitary operators. Therefore, the scalar product is invariant under these
transformations and all transition probabilities are the same for all
observers connected by Poincar\'e transformations. Thus, in both coordinate
and momentum representations, wave mechanics of photons is a fully
relativistic theory.

\section{Localizability of photons\label{localizability}}

The problem of localization of relativistic systems has been first posed and
solved by \cite{NW_49} and later refined by \cite{Wightman_62}. According to
this analysis it is possible to define position operators and localized
states for massive particles and for massless particles of spin 0, but not
for massless particles with spin. Thus, the position operator in the sense
of Newton and Wigner does not exist for photons (cf. also a recent tutorial
review on that subject by \cite{RS_92}). As a simple heuristic explanation,
why position operator for the photon does not exist, one may observe
(\cite{Pryce_48}) that the multiplication by ${\bf r}$ can not be applied to
the photon wave function because it destroys the divergence condition
(\ref{cmax2}).

A weaker definition of localization that is applicable even when the
position operator does not exist, was proposed by \cite{JP_67} and a very
detailed analysis of this problem has been given by \cite{Amrein_69}. The
Jauch-Piron localizability allows for noncompatibility of "photon position
measurements" in overlapping regions. The main weakness of such an abstract
analysis is that an operational definition of the photon position
measurement for photons has not been incorporated into it. The existence of
position measurements for photons is just taken for granted regardless of
the feasibility of their physical realizations. When a realistic model of
the photon detector is brought in, it is the wave function $\Psi$ rather
than $\Phi$ that appears as the correct probability amplitude for
photodetection (\cite{MW_95}). Thus, in practical applications the energy
wave functions $\Psi$ always seem to play a dominant role.

It must, however, be stressed that even for massive particles, the
localization is not perfect, because it is not relativistically invariant.
Two observers who are in relative motion would not quite agree as to the
localization region of a relativistic particle. This follows from the fact
that the Newton-Wigner wave function $\psi^{NW}$ is related to the
relativistic wave function $\psi$ that transforms locally under Poincar\'e
transformations by a nonlocal transformation (cf. \cite{Haag_93})
\begin{eqnarray}
 \psi^{NW}({\bf r})
 = \int\!d^3r'K({\bf r - r'})\psi({\bf r'}),\label{nw_loc}
\end{eqnarray}
where the kernel $K$ can be represented in terms of the Macdonald function
\begin{eqnarray}
K({\bf r}) =
\sqrt{\frac{\pi}{2}}(\frac{2mc}{r\hbar})^{5/4}K_{5/4}(mcr/\hbar).
\end{eqnarray}
In the limit, when $m \to 0$, $K({\bf r}) \to \pi/(2\pi r)^{-5/2}$ and
  (\ref{nw_loc}) becomes the relation (\ref{lp_loc}) between the local wave
function of the photon and the Landau-Peierls wave function. Thus, the
difference between the localizability of massive particles and photons is
not that great. In both cases, localization can not be defined in a
relativistic manner. However, for massive particles departures from strict
localization are only exponentially small due to the fast decay of the
Macdonald function in eq. (\ref{nw_loc}). In the nonrelativistic limit,
when $c \to \infty$, the exponential tails become infinitely sharp and the localization is restored.

Difficulties with the position operator for relativistic particles have a
profound origin connected with the structure of the Poincar\'e group. In
nonrelativistic physics the position operator is the generator (up to a
factor of mass) of Galilean transformations (cf., for example,
\cite{Gottfried_66}, \cite{Weinberg_95}, p. 62). In a relativistic theory,
the Galilean transformations are replaced by the Lorentz transformations and
the position operator (multiplied by the mass) is replaced by the boost
generator ${\bf K}$. The main difference between Galilean and Lorentz
transformation affecting the discussion of localizability is that boost
generators {\em do not commute}. Therefore, one may only hope to localize
relativistic particles in one direction at a time. The possibility to
localize photons in one direction has been discussed in general terms as the
``front'' description by \cite{AS_60}. The eigenfunctions of the boost
operator $K_z$ given in \S \ref{eigenvalue} may serve as an explicit
realization of the front description for the photon.

The considerations of photon localizability, while important for the
understanding of some fundamental issues, do not influence much the
practical applications of the photon wave function. All that really should
matter there is that the wave function be precisely defined and that its
interpretation be not extended beyond the limits of applicability.

\section{Phase-space description of a photon\label{phase-space}}

Distribution functions in phase space are a very convenient tool in the
description of statistical properties and the study of the classical limit
of wave mechanics. A direct analog of the Wigner function (\cite{Wigner_32})
introduced in wave mechanics may also be introduced for photons with the
help of the photon wave function. This is done by Fourier transforming the
product of the wave function and its complex conjugate. Fourier transforms
of the electromagnetic fields similar to the Wigner function have been
introduced in optics, first by \cite{Walther_68} in the two-dimensional
context of radiative transfer theory and then by \cite{Wolf_76} and by
Sudarshan [1979, 1981a,b] in the three-dimensional case. In these papers
phase-space distribution functions were defined for the {\em stationary states} of the electromagnetic field only and they were treated as functions of the frequency. The time-dependent distribution functions can be defined
(\cite{IBB_94}) with the use of the time-dependent wave function. The only
formal difference between the standard definition of the Wigner function in
nonrelativistic wave mechanics of massive particles and the case of photons
is the presence of vector indices. Thus, the photon distribution function
in phase space is not a single scalar function but rather a $6\times 6$
Hermitian matrix defined as follows
\begin{eqnarray}
W_{ab}({\bf r}, {\bf k},t) = \int\! d^3s\; e^{-i{\bf
k}\cdot{\bf s}}\; \Psi_a({\bf r} + {\bf s}/2,t) \Psi_b^*({\bf r} - {\bf
s}/2,t).\label{w}
\end{eqnarray}
Similar multi-component distribution functions arise also for a Dirac
particle and one can use some of the techniques developed by \cite{BGR_91}
to deal with such functions.

Every $6\times 6$ Hermitian matrix can be written in the following block form
\begin{eqnarray}
 W_{ab} = \left[ \begin{array}{cc}
 W^0_{ij}+W^3_{ij} & W^1_{ij} - iW^2_{ij}\\~ & ~\\
 W^1_{ij} + iW^2_{ij} & W^0_{ij}-W^3_{ij}
 \end{array}\right],
\end{eqnarray}
where all $3\times 3$ matrices $W^\alpha_{ij}$ are Hermitian. This
decomposition can also be expressed in terms of the $\rho$ matrices
\begin{eqnarray}
 W_{ab} = \rho_0 W^0_{ij} + \rho_1 W^1_{ij}
 + \rho_2 W^2_{ij} + \rho_3  W^3_{ij}.
\end{eqnarray}
The vector indices $i$ and $j$ refer to the components within the upper and
lower parts of the wave function and the matrices $\rho$ act on these parts
as a whole. The most general photon distribution function, as seen from this
analysis, is quite complicated. In general, when the medium induces mixing
of the two polarization states, all components of the distribution function
are needed. However, when photons propagate in free space, only a subset of
these components is sufficient to account for the dynamical properties of
photon beams. The simplest case is that of a given helicity. A more
interesting case is that of an unpolarized photon beam: a mixture of both
helicities with equal weights. This state has to be described by the
distribution function because a mixed state can not be treated by pure
Maxwell's theory. In all these cases phase-space dynamics can be described
by a $3\times 3$ Hermitian matrix, i.e., by just one scalar function and one
vector function. To this end, one may introduce the following reduced
distribution function
\begin{eqnarray}
W_{ij}({\bf r}, {\bf k},t) = \int\! d^3s\; e^{-i{\bf
k}\cdot{\bf s}}\; \psi_i({\bf r} + {\bf s}/2,t) \psi_j^*({\bf r} - {\bf
s}/2,t),\label{wig}
\end{eqnarray}
where $\psi_i$ are the upper or the lower components of the original wave
function. The Hermitian matrix $W^\alpha_{ij}$ can be decomposed into a real
symmetric tensor and a real vector according to the formula

\begin{eqnarray}
W^\alpha_{ij} = w^\alpha_{ij} + \frac{c}{2i}\epsilon_{ijk} u^\alpha_k.
\end{eqnarray}
The tensor corresponds to the symmetric part of $W^\alpha_{ij}$ and the vector corresponds to the antisymmetric part. The factor of $c$ has been separated out in the second term since the vector ${\bf u}$ is related to the momentum density, while the trace of $w_{ij}$ is related to the energy density (\cite{IBB_94}).

The equations satisfied by the components of the photon distribution
function in free space can be obtained from Maxwell's equations
(\ref{cmax1}) and (\ref{cmax2}) for the vector ${\bf F}$,
\begin{eqnarray}
\partial_t W_{ij} = c\Big({\bf k} + \frac{i}{2}{\bf
\nabla}\Big)_m \epsilon_{mil}\,W_{lj} - c\Big({\bf k} -
\frac{i}{2}{\bf \nabla}\Big)_m W_{il}\,\epsilon_{mlj},\\
\Big({\bf k} + \frac{i}{2}{\bf \nabla}\Big)_i\,W_{ij} = 0 = \Big({\bf k} -
\frac{i}{2}{\bf \nabla}\Big)_j\,W_{ij}.\hspace{25pt}
\end{eqnarray}
This leads to the following set of coupled evolution equations for the real
components $w_{ij}$ and $u_i$
\label{evl}
\begin{eqnarray}
 \partial_t w_{ij} = -c\epsilon_{ilk} k_l w_{kj}
 -c\epsilon_{jlk} k_l w_{ki} - \frac{c^2}{2}(\nabla_i u_j + \nabla_j u_i
 -\delta_{ij} \nabla_k u_k),\\
 \partial_t u_i = -c\epsilon_{ijk} k_j u_k - \frac{1}{2}
 (\nabla_j w_{ij} - \nabla_i w_{jj}),\hspace{100pt}
\end{eqnarray}
and to the subsidiary conditions
\label{sub}
\begin{eqnarray}
 c\epsilon_{ijk} k_j u_k = \nabla_j w_{ij},\label{sub1}\\
 c\epsilon_{ijk} \nabla_j u_k = 4k_j w_{ij}.\label{sub2}
\end{eqnarray}
The ${\bf k}$-dependent terms in the evolution equations describe a uniform
rotation of the vector $u_i$ and of the tensor $w_{ij}$ around the wave vector ${\bf k}$ so that these terms can be eliminated by ``going to a rotating coordinate frame''.

With the help of the subsidiary conditions (\ref{sub}) one can eliminate the
remaining components and obtain from the evolution equations (\ref{evl})
the equations for $w=\sum w_{ii}$ and ${\bf u}$
\begin{eqnarray}
\partial_t w &=& -c^2\nabla_i u_,\\
\partial_t u_i &=& -2c\epsilon_{ijk} k_j u_k - \nabla_i w.
\end{eqnarray}
These evolution equations do form a simple, self-contained set. However, as
it is always the case with the phase-space distribution functions in wave
mechanics, not all solutions of these equations are admissible. Only those
distribution functions are allowed that can be represented in the form
(\ref{wig}) at the initial time (with ${\bf u}$ and $w$ satisfying the
subsidiary conditions (\ref{sub1}) and (\ref{sub2})).

\section{Hydrodynamic formulation\label{hydrodynamic}}

It has been shown by \cite{Madelung_26} that the Schr\"odinger wave equation
can be cast into a hydrodynamic form. In this form, the complex wave
function is replaced by real variables: the probability density $\rho$ and
the velocity ${\bf u}$ of the probability flow. The wave equation is
replaced by the hydrodynamic evolution equations for the variables $\rho$
and ${\bf u}$. In order to reduce the number of functions from four to the
original two, one has to impose auxiliary conditions --- the quantization
condition --- on the velocity field. Later, other wave equations in quantum
mechanics (Pauli, Dirac, Weyl) were also presented in the hydrodynamic form.
The wave equation for the photon wave function is not an exception in this
respect. It can also be written (\cite{IBB_96b}) as a set of equations for
real hydrodynamic-like variables. Since the Riemann-Silberstein vector ${\bf
F}$ carries all the information about the photon wave function, one may use
${\bf F}$ to define these variables. They comprise the energy density $\rho$
and the velocity of the energy flow ${\bf v}$,
\begin{eqnarray}
 \rho({\bf r},t) = {\bf F}^*({\bf r},t)\cdot{\bf F}({\bf r},t),\;\;
 \rho({\bf r},t){\bf v}({\bf r},t)
 = \frac{c}{2i}{\bf F}^*({\bf r},t)\times{\bf F}({\bf r},t),
\end{eqnarray}

the components $t_{ij}$ of the following tensor
\begin{eqnarray}
 t_{ij}({\bf r},t) = F^*_i({\bf r},t)F_j({\bf r},t)
 + F^*_j({\bf r},t)F_i({\bf r},t),
\end{eqnarray}
and another vector ${\bf u}$,
\begin{eqnarray}
 \rho({\bf r},t){\bf u}({\bf r},t)
 = \frac{c}{2i}\bigl({\bf F}^*({\bf r},t)\nabla{\bf F}({\bf r},t)
 -(\nabla{\bf F}^*({\bf r},t)){\bf F}({\bf r},t)\bigr).
\end{eqnarray}
Owing to the existence of the following identities satisfied by the
hydrodynamic variables
\begin{eqnarray}
  t_{ii} = 2c,\;\;\;\;
  v_i t_{ik} = 0,\;\;\;\;
  t_{ij} t_{ij} = 4c^2 - 2\vec v^2,
\end{eqnarray}
only one component of $t_{ij}$ is arbitrary, but the hydrodynamic
equations look more symmetric when all the components are treated on equal
footing. The number of algebraically independent hydrodynamic variables is
reduced from eight to six (the number of degrees of freedom described by
${\bf F}$) by the following quantization condition
\begin{equation}
\int\!d\vec S\!\cdot\!\bigl[\nabla\times\vec u
 - \frac{1}{8c^3}\varepsilon_{ijk}
  (v_i{\bf\nabla}v_j\times{\bf\nabla}v_k +
  v_i{\bf\nabla}t_{jl}\times{\bf\nabla}t_{kl} -
  2t_{il}{\bf\nabla}t_{jl}\times{\bf\nabla}v_k)\bigr] = 2\pi n,
\end{equation}
where $n$ is a natural number. This condition must hold for every choice of
the integration surface and it states, in essence, that the phase of the
wave function is uniquely defined (up to an overall constant phase).

The evolution equations for the hydrodynamic variables are
\begin{eqnarray}
  \partial_t\rho + (\vec v\!\cdot\!{\bf\nabla})\rho &=&
  -\rho({\bf\nabla}\!\cdot\!\vec v),\\
  \partial_t v_i + (\vec v\!\cdot\!{\bf\nabla}) v_i &=&
  \frac{1}{\rho}\partial_{j}(-c^2\rho\delta_{ij} +
  \rho v_i v_j +\rho t_{ij}),\\
  \partial_t t_{ij} + (\vec v\!\cdot\!{\bf\nabla}) t_{ij} &=&
  \frac{1}{\rho}(t_{ij}v_k\partial_{k}\rho-
  c\delta_{ij}v_k\partial_{k}\rho+
  \frac{c}{2}(v_i\partial_{j}+v_j\partial_{i})\rho)\nonumber\\
  &+&\delta_{ij}v_k\partial_{l} t_{kl}+2v_k\partial_{k} t_{ij}+
  c\varepsilon_{ikl}u_k t_{lj}+c\varepsilon_{jkl}u_k t_{li}\nonumber\\
  &+&(v_k\partial_{k} t_{ij}-t_{ij}\partial_{k} v_k)+
  \frac{1}{2}(t_{ik}\partial_{k} v_j+t_{jk}\partial_{k} v_i)\nonumber\\
  &-&\frac{1}{2}(v_i\partial_{k} t_{kj}+v_j\partial_{k} t_{ki}+
  v_k\partial_{i} t_{kj}+v_k\partial_{j} t_{ki}),\\
  \partial_t u_i + (\vec v\!\cdot\!{\bf\nabla}) u_i &=&
  \frac{1}{4c\rho}\partial_{j}\bigl[\rho\varepsilon_{jkl}
  (t_{km}\partial_{i} t_{ml} +v_k\partial_{i} v_l)\bigr].
\end{eqnarray}
They must be supplemented by the equations that express the divergence
condition (\ref{cmax2})
\begin{eqnarray}
  &&\frac{1}{2}\partial_{k}(\rho t_{ik})+\rho \varepsilon_{ijk} v_j u_k\\
 &+& \frac{\rho}{4c}(t_{jk}\partial_{k} t_{ij}-t_{ij}\partial_{k} t_{jk}+
  v_k\partial_{k} v_i-v_i\partial_{k} v_k) = 0,\nonumber\\
  &&\frac{1}{2}\partial_{k}(\rho\varepsilon_{ikl}v_l)+\rho t_{ik}u_k\\
 &+& \frac{\rho}{4c}\bigl[\varepsilon_{jkl}(t_{il}\partial_{k} v_j-
  v_j\partial_{k} t_{il})+
  \varepsilon_{ijl}(t_{kl}\partial_{k} v_j-v_j
  \partial_{k} t_{kl})\bigr]=0.\nonumber
\end{eqnarray}

Hydrodynamic description of the photon dynamics is not simple but its
existence underscores again the unification of the quantum theory of the
photon with the rest of quantum mechanics. Everything that can be done with
other particles, can be also done for the photons.

\section[Wave function in curved space]{Photon wave function in
non-Cartesian coordinate systems and in curved space\label{curved}}

It has been shown by \cite{Skrotskii_57} and \cite{Plebanski_60} (for a
pedagogical review, see \cite{SS_84}) that the propagation of the
electromagnetic field in arbitrary coordinate systems including also the
case of curved spacetime may be described by Maxwell's equations, with all
the information about the spacetime geometry contained in the relations
connecting the field vectors ${\bf E,B}$ and ${\bf D,H}$. This discovery can
be further enhanced by the observation (\cite{IBB_94}) that in
contradistinction to the case of an inhomogeneous medium, in the
gravitational field the two photon helicities do not mix. This follows from
the fact that for arbitrary metric $g_{\mu\nu}$ the constitutive relations
can be written as a single equation connecting two complex vectors: the
vector ${\bf F}({\bf r},t)$ defined by the formula (\ref{rsv0}) and a new
vector ${\bf G}({\bf r},t)$ defined as
\begin{eqnarray}
 {\bf G}({\bf r}, t) = \frac{1}{\sqrt{2}}\left(\frac{{\bf E}({\bf r},t)}
 {\sqrt{\mu_0}} + i\frac{{\bf H}({\bf r},t)}
 {\sqrt{\epsilon_0}}\right).\label{vec_g}
\end{eqnarray}
In curved space (or in curvilinear coordinates), the constitutive relations
for two complex vectors ${\bf F}({\bf r},t)$ and ${\bf G}({\bf r},t)$ have
the form
\begin{eqnarray}
 F^i = -\frac{1}{g_{00}} \bigl(\sqrt{-g}g^{ij}
 + ig_{0k}\varepsilon^{ikj}\bigr)G_j,\label{cr1}\\
 G_i = -\frac{1}{g^{00}} \bigl(g_{ij}/\sqrt{-g}
 - ig^{0k}\varepsilon_{ikj}\bigr)F^j,\label{cr2}
\end{eqnarray}
where $g_{\mu \nu}$ is the metric tensor, $g^{\mu \nu}$ is its inverse, and $g$ is the
determinant of $g_{\mu \nu}$. The Maxwell equations expressed in terms of vectors
${\bf G}$ and ${\bf F}$ in curved spacetime are the same as in flat space
\begin{eqnarray}
 i\partial_t{\bf F}({\bf r},t) &=& {\bf \nabla}\times{\bf G}({\bf r},t),\\
 {\bf \nabla}\!\cdot\!{\bf F}({\bf r},t) &=& 0,
\end{eqnarray}
In these equations, all derivatives are {\it ordinary} (not covariant)
derivatives as in flat space. The whole difference is in the form of the
constitutive relations (\ref{cr1}) and (\ref{cr2}). These relations contain
all information about the gravitational field or the curvilinear coordinate
system. Since the relations between ${\bf G}$ and ${\bf F}$ are linear, one
 may write again two separate wave equations for the two helicity states as
in flat space. By combining these two equations, one obtains the following
wave equation for the six-component photon wave function ${\cal F}({\bf
r},t)$
\begin{eqnarray}
i\partial_t{\cal F}({\bf r},t) = \rho_3 {\bf \nabla}\times{\cal G}({\bf
r},t),
\end{eqnarray}
where
\begin{eqnarray}
{\cal G}_i = -\frac{1}{g^{00}} \bigl(g_{ij}/\sqrt{-g} - i\rho_3
g^{0k}\varepsilon_{ikj}\bigr){\cal F}^j.
\end{eqnarray}
These equations contain only the matrix ${\rho_3}$ that does not mix the
helicity states.

The true photon wave function $\Psi({\bf r},t)$ may be introduced only in
the time-independent case, when the separation of ${\cal F}({\bf r},t)$ into
positive and negative energy parts is well defined.

\section[Wave function as a spinor]{Photon wave function as a spinor
field\label{spinor}}

Soon after the formulation of spinor calculus by van der Waerden in the
context of the Dirac equation, it has been discovered by \cite{LU_31} that
the Maxwell equations can also be cast into a spinorial form. The spinor
representation of the electromagnetic field and the Riemann-Silberstein
vector are closely connected. The components of the vector ${\bf F}$ are
related to the components of a second rank symmetric spinor $\phi_{AB}$
\begin{eqnarray}
\phi_{00} &=& -F_x + iF_y,\\
\phi_{01} &=& F_z,\\
\phi_{11} &=& F_x + iF_y,
\end{eqnarray}
and the components of the complex conjugate vector ${\bf F^*}$ are related
to the components of a second rank symmetric primed spinor $\phi^{A'B'}$,
\begin{eqnarray}
\phi^{0'0'} &=& -F^*_x - iF^*_y,\\
\phi^{0'1'} &=& -F^*_z,\\
\phi^{1'1'} &=& F^*_x - iF^*_y.
\end{eqnarray}
The property that even in curved space both helicities
propagate without mixing is in the spinorial formalism a simple consequence of the fact that both spinors $\phi_{AB}$ and $\phi^{A'B'}$ satisfy separate wave equations
(\cite{LU_31}, \cite{PR_84})
\begin{eqnarray}
 \sigma^{\mu C'A}\partial_\mu\phi_{AB}({\bf r},t)
 &=& 0,\label{max_spin1}\\
 \sigma^\mu_{\;\;CA'}\partial_\mu\phi^{A'B'}({\bf r},t)
 &=& 0,\label{max_spin2}
\end{eqnarray}
where the matrices $\sigma^{\mu A'B}$ and $\sigma^\mu_{\;\;A'B}$ are built
from the unit matrix and the Pauli matrices,
\begin{eqnarray}
 (\sigma^{\mu A'B}) &=& (I, \mbox{\boldmath$\sigma$})^{A'B},\\
 (\sigma^\mu_{\;\;AB'}) &=& (I, -\mbox{\boldmath$\sigma$})_{AB'}.
\end{eqnarray}
Under a Lorentz transformation the second rank spinor changes according to
the formula
\begin{eqnarray}
 \phi_{AB}'({\bf r'},t')
 = S_A^{\;\;C}S_B^{\;\;D}\phi_{CD}({\bf r},t),
\end{eqnarray}
where $S_A^{\;\;C}$ is a $2\times 2$ matrix of the fundamental (spinor)
representation of the Lorentz group. Thus, from the point of view of the
representation theory of the Lorentz and Poincar\'e groups (\cite{SW_78},
\cite{Weinberg_95}, p. 231), the photon wave functions for a given helicity
are just the three-component fields that transform as irreducible
representations $(1,0)$ or $(0,1)$ of the proper Lorentz group (without
reflections). In order to accommodate reflections one must combine both
representations and introduce the six-dimensional objects ${\cal F}$ or
$\Psi$.

Eqs. (\ref{max_spin1}) (and similarly eqs. (\ref{max_spin2})) represent a
set of four equations obeyed by four components $(\phi_{11}, \phi_{12},
\phi_{21}, \phi_{22})$ of the second order spinor. All four equations can be
written in the form of the Dirac equation (\cite{Ohmura_56},
\cite{Moses_59})
\begin{eqnarray}
 i\hbar\partial_t\phi({\bf r},t)
 = \mbox{\boldmath$\alpha$}\!\cdot\!\frac{\hbar}{i}{\nabla}\phi({\bf r},t),
\end{eqnarray}
where the matrices $\mbox{\boldmath$\alpha$}$ are
\begin{eqnarray}{
\arraycolsep=2pt
 \alpha_x = \left[ \begin{array}{cccc}
 0 & 0 & 1 & 0\\
 0 & 0 & 0 & 1\\
 1 & 0 & 0 & 0\\
 0 & 1 & 0 & 0\end{array}
 \right]\!\!,\;\;
 \alpha_y = \left[ \begin{array}{cccc}
 0 & 0 & -i & 0\\
 0 & 0 & 0 & -i\\
 i & 0 & 0 & 0\\
 0 & i & 0 & 0\end{array}
 \right]\!\!,\;\;
 \alpha_z = \left[ \begin{array}{cccc}
 1 & 0 & 0 & 0\\
 0 & 1 & 0 & 0\\
 0 & 0 & -1 & 0\\
 0 & 0 & 0 & -1\end{array}
 \right]\!\!.}\label{alpha_mat}
\end{eqnarray}
One may check that in this formulation the divergence condition takes on the
following simple algebraic form: $\phi_{12} = \phi_{21}$.

The Maxwell equations expressed in spinor notation and the Weyl equation
provide just the simplest examples from a hierarchy of wave equations for
massless fields described by symmetric spinors $\phi_{B_1 B_2 \cdots B_n}$
or $\phi^{B'_1 B'_2 \cdots B'_n}$. All these equations have the form
(\cite{PR_84})
\begin{eqnarray}
\sigma^{\mu C'A}\nabla_\mu\phi_{AB_1 B_2 \cdots B_{n-1}}({\bf r},t) = 0,\\
\sigma^\mu_{\;\;CA'}\nabla_\mu\phi^{A'B'_1 B'_2 \cdots B'_{n-1}}({\bf r},t)
 = 0.
\end{eqnarray}
This universality of massless wave equations for all spins gives an
additional argument for treating the Riemann-Silberstein vector as the
photon wave function.

\section[WAVE FUNCTIONS AND MODE EXPANSION]{Photon wave functions and mode
expansion of the electromagnetic field\label{mode}}

The concept of the photon wave function is also useful in the process of
quantization of the electromagnetic field. One may simply apply the
procedure of second quantization to the photon wave function in the same
manner as one does it for other field operators. In order to see this
analogy, one may recall that the field operator ${\hat\psi}({\bf r})$ for,
say the electron field, is built from a complete set of wave functions for
the electrons $\psi^+_n({\bf r})$ and from a complete set of wave functions
for positrons $\psi^-_n({\bf r})$ according to the following rule (cf., for
example, \cite{Schweber_61, BBBB_75, Weinberg_95}
\begin{eqnarray}
 {\hat\psi}({\bf r},t) = \sum_n(\psi^+_n({\bf r},t){\hat a}_n
 + \psi^-_n({\bf r},t){\hat b}^{\dagger}_n),
\end{eqnarray}
where ${\hat a}_n$ and ${\hat b}_n$ are the annihilation operators for
electrons and positrons respectively. The second part of the field operator
is related to the first by the operation of charge conjugation performed on
the wave functions and on the operators. The analog of charge conjugation
for photon wave functions is given by eq. (\ref{conjugation}). Following
this procedure, one may construct the field operator of the electromagnetic
field in the form
\begin{eqnarray}
 {\hat{\cal F}}({\bf r},t) = \sum_n(\Psi_n({\bf r}, t){\hat c}_n
 + \rho_1\Psi^*_n({\bf r},t){\hat c}^{\dagger}_n),
 \label{field_op}
\end{eqnarray}
where the identity of particles and antiparticles for photons has been taken
into account by using only one set of creation and annihilation operators.
The field operator (\ref{field_op}) is non-Hermitian but it satisfies the
particle-antiparticle conjugation condition (\ref{conjugation}). Therefore,
it only has six Hermitian components. These Hermitian operators are
identified as the field operators ${\hat{\bf D}}({\bf r},t)$ and ${\hat{\bf
B}}({\bf r},t)$ and they are obtained from ${\hat{\cal F}}$ through the
formula
\begin{eqnarray}
 {{\hat{\bf D}}\brack{\hat{\bf B}}}
 = \frac{1}{2}{1\;\;\;1\brack-i\;\;i}{\hat{\cal F}}.
\end{eqnarray}

As a direct consequence of the formula (\ref{field_op}) one may identify the
photon wave functions in the second-quantized theory with the matrix
elements of the electromagnetic field operators ${\hat{\cal F}}$ or
${\hat{\bf D}}$ and ${\hat{\bf B}}$ taken between one-particle states and
the vacuum
\begin{eqnarray}
 \Psi_n({\bf r}, t)
 = \langle 0\vert{\hat{\cal F}}({\bf r},t)
 {\hat c}^{\dagger}_n\vert 0\rangle.
\end{eqnarray}

In the simplest case of the free field, when the complete set of photon wave
functions may be labeled by the wave vector ${\bf k}$ and helicity
$\lambda$, the formula (\ref{field_op}) takes on the form
\begin{eqnarray}
 {\hat{\cal F}}({\bf r}, t) \!=\!
 \sqrt{\hbar c}\!\!\!\int\!\!\!\frac{d^3k}{(2\pi)^3}\!
 {{\bf e}({\bf k},1)({\hat c}({\bf k},1)
 e^{-i\omega t + i{\bf k}\cdot{\bf r}}
 \!+\! {\hat c}^{\dagger}\!({\bf k},-1)
 e^{i\omega t - i{\bf k}\cdot{\bf r}})
 \brack
 {\bf e}({\bf k},-1)({\hat c}({\bf k},-1)
 e^{-i\omega t + i{\bf k}\cdot{\bf r}}
 \!+\! {\hat c}^{\dagger}\!({\bf k},1)
 e^{i\omega t - i{\bf k}\cdot{\bf r}})}\!.
 \label{emf_op}
\end{eqnarray}
In the presence of a medium, the expansion (\ref{field_op}) of the
electromagnetic field operator requires the knowledge of a complete set of
wave functions $\Psi_n$ that satisfy the photon wave equation in the medium.
These functions are called usually the mode functions of the electromagnetic
field (cf., for example, \cite{Louisell_73} p. 240 and \cite{MW_95} p. 905)
but the term photon wave functions is perhaps more appropriate
(\cite{Moses_73, BBBB_75}. The advantage of using the terminology of wave
functions is that it brings in all the associations with wave mechanics and
makes the classification of the modes more transparent. In particular, one
may use the quantum-mechanical notion of eigenfunctions and eigenvalues to
classify the functions used in the mode expansion (\cite{Moses_73}) and also
borrow from quantum mechanics the methods of proving their completeness
(\cite{BBB_72}).

This discussion shows that the photon wave function is not restricted to the
{\em wave mechanics} of photons. The same wave functions also appear as the mode functions in the expansion of the electromagnetic field operators.

\section{Summary}

The aim of this review was to collect and explain all basic properties of a
certain well defined mathematical object --- a six-component function of
space-time variables --- that describes the quantum state of the photon.
Whether one decides to call this object the photon wave function in
coordinate representation is a matter of opinion since some properties known
from wave mechanics of massive particles are missing. The most essential
property that does not hold for the photon wave function is that the
argument ${\bf r}$ of the wave function can not be directly associated with
the position operator of the photon. The position operator for the photon
simply {\em does not exist}. However, one should remember that also for massive particles the true position operator exists only in the nonrelativistic approximation. The concept of localization associated with the Newton-Wigner position operator is {\em not relativistically invariant}. Since photons can not be described in a nonrelativistic manner, there is no approximate position operator.

The strongest argument that can be made for the photon wave function in
coordinate representation is based on the most fundamental property of
quantum states --- on the {\em principle of superposition}. According to the
superposition principle, wave functions form a linear space. By adding wave
functions one obtains again legitimate wave function. Once this principle is
accepted, the existence of photon wave functions in coordinate
representation follows from the existence of the photon wave functions in
momentum representation and these functions are genuine by all standards;
their existence simply follows from relativistic quantum kinematics (or more
precisely from the representation theory of the Poincar\'e group). The
Fourier integral (\ref{gen_psi}) represents a special combination of
momentum space wave functions with different momenta and as a matter of
principle, such linear combinations are certainly allowed. One may only
argue which superpositions to take as more natural or useful but totally
rejecting the very concept of the photon wave function in coordinate
representation is tantamount to rejecting the superposition principle
altogether.

There is not much advantage in using the photon wave function in coordinate
representation to perform calculations for photons moving in {\em free space}. The relation of this wave function to momentum wave function is so
straightforward that one may as well stick to momentum representation. It is
only in the presence of a medium, especially in an inhomogeneous medium,
that the photon wave function in coordinate representation becomes useful and even essential. Only in the coordinate representation one may hope to solve the eigenvalue problems and to take into account the boundary conditions.

The introduction of the wave function for the photon has many significant
benefits. The photon wave function enables one to formulate a consistent
wave mechanics of photons that could be often used as a convenient tool in
the quantum description of electromagnetic fields, {\em independently} of
the formalism of second quantization. In other words, in constructing
quantum theories of photons one may proceed, as in quantum theory of all
other particles, through two stages. At the first stage one introduces wave
functions and a wave equation obeyed by these wave functions. At the second
stage one upgrades the wave functions to the level of field operators in
order to deal more effectively with the states involving many
indistinguishable particles and to allow for processes in which the number
of particles is not conserved. Many methods that have proven very useful in
the study of particles described by the Schr\"odinger wave functions can
also be implemented for photons leading to some new insights. These methods
include relationships between symmetries and operators, the definitions of
various sets of modes for the electromagnetic field and their completeness
relations, eigenvalue problems for various observables(\S \ref{eigenvalue}),
phase-space representation (\S \ref{phase-space}), and
the hydrodynamic formulation (\S \ref{hydrodynamic}). Finally, there are
important logical and pedagogical advantages coming from the use of the
photon wave function. The quantum mechanical description of {\em all}
particles, including photons, becomes uniform.

\vspace{.5cm}
\centerline{\bf Acknowledgments}
\vspace{.5cm}

Four lectures based on this review were presented while I was a Senior
Visiting Fellow at the Rochester Theory Center for Optical Science and
Engineering in February 1996. I would also like to acknowledge discussions
with Z. Bialynicka-Birula, M. Czachor, J. H. Eberly, A. Or\l owski, W.
Schleich, M. O. Scully, and E. Wolf.

\end{document}